# The Exchange Hole in the Dirac Sea

F. J. Himpsel

Physics Department, University of Wisconsin-Madison, Madison WI 53706, USA
fhimpsel@wisc.edu

**Abstract**

This work is motivated by the long-standing question about the internal stability of the electron. While one cannot investigate internal properties of a point-like particle, it is fair to analyze the response of the Dirac sea to an electron. For this purpose the concept of an exchange hole is generalized from the Fermi sea of condensed matter physics to the Dirac sea of quantum electrodynamics. In order to conserve charge and angular momentum, an exchange electron is added to the exchange hole, forming a neutral exchange exciton. Consequently, the pair correlation defining the exchange hole is generalized to a three-fermion correlation. An approximation by sequential pair correlations provides analytic results for charge densities, average distances, and force densities of the exchange hole and electron. The analysis is supported by analogies to the negative ion of positronium.

**Contents**

## 1. Introduction

The internal stability of the electron has remained an unsolved problem for more than a century [1]. In classical electrodynamics, a charge cloud would Coulomb-explode due to internal electrostatic repulsion. This problem disappears in quantum electrodynamics because of the exclusion principle. Self-interaction involves the same electron twice at the same location, once as the source of an electric field and a second time as the particle that interacts with this field. More quantitatively, a many-electron wave function has to be antisymmetric with respect to exchanging electrons. The corresponding Slater determinant vanishes if it contains the same electron twice. In Hartree-Fock theory the self-Coulomb interaction in the Hartree term is canceled by the self-exchange in the Fock term. Since the exchange interaction plays such a crucial role in preventing the classical Coulomb explosion, it becomes a prime candidate for solving the stability problem.

In quantum electrodynamics the direct self-interaction of an electron is replaced by indirect interactions via the Dirac sea of vacuum electrons and positrons. For example, an electron can polarize the Dirac sea. The resulting vacuum polarization then generates an electric field that acts back on the electron. This indirect Coulomb self-interaction is overshadowed by the corresponding exchange interaction. It entangles the extra electron with the vacuum electrons without being assisted by an electric field.

The effect of the exchange interaction on the electron self-energy was already investigated in the early years of quantum electrodynamics by Weisskopf [2]. He calculated the pair correlation between an electron added to the vacuum and the Dirac sea. The result was interpreted as a point-like hole at the location of the extra electron, surrounded by a spread-out electron. The hole and the extra electron were assumed to compensate each other, leaving the spread-out electron. Thereby the classical 1/r divergence in the Coulomb self-energy of a point charge weakens to a logarithm. The latter is eliminated by mass renormalization in modern quantum electrodynamics.

Weisskopf's work is revisited here from the perspective of condensed matter physics, where the Dirac sea is replaced by the Fermi sea, i.e., an electron gas neutralized by a positive background charge. For this system the concept of an exchange hole was developed [3]-[7]. It is derived from the pair correlation between electrons in the Fermi sea. More intuitively one can argue that the exclusion principle pushes other electrons with the same spin away from a reference electron, leaving a positive charge and an opposite spin. Those compensate the charge and spin of the reference electron. Since this occurs over atomic distances, the electron-electron repulsion is reduced dramatically. Such short-range screening validates the simple Fermi liquid model of metals which assumes nearly-free electrons.

While Weisskopf's pioneering work on the electron self-interaction is a remarkable achievement that has stood the test of time, questions about the sign of the induced charge suggest a re-analysis from the perspective of condensed matter physics. With the definition given here, Weisskopf's spread-out electron becomes a spread-out hole, like the exchange hole in the Fermi sea. This is due to our choice of a reference electron inside the Dirac sea (versus an extra electron) – analogous to the definition of the exchange hole in the Fermi sea. The point-like hole in Weisskopf's analysis was defined from states with positive energy only. In our analysis such δ-functions appear as well, but they cancel each other after combining contributions from electrons and positrons in the Dirac sea.

When transferring the concept of the exchange hole from the Fermi sea to the Dirac sea we take into account relativistic invariance, particle-antiparticle symmetry, and charge conservation. The result is again an exchange hole, but the hole is surrounded by an exchange electron. The latter can be identified with the electron displaced by the exclusion principle. It may also be viewed as the "exchange hole of the exchange hole", i.e., the response of the Dirac sea to the exchange hole. The transition from hole to electron character occurs near the reduced Compton wavelength $\lambda_C = 1/m_e$ (in units of $\hbar, c$). This length is 2-3 orders of magnitude shorter than screening lengths in metals.

In contrast to the metallic Fermi sea the overall response of the Dirac sea has to remain neutral due to charge conservation. Likewise, the angular momentum needs to be conserved. These requirements lead to the addition of the exchange electron which has the same charge and spin as the original electron and thereby compensates those of the exchange hole. Exchange electron and hole form an exchange exciton in the spin singlet state. The appearance of the exchange electron can be traced to the insulating character of the Dirac sea. While the exchange electron can be removed to infinity in a metal with an infinitesimally-small amount of energy, an electron cannot be moved in the Dirac sea without creating an electron-positron pair. That would require the energy $2m_e$.

The ultimate goal of this work is a force density balance as criterion for the stability of the electron itself. In prototypical electron systems studied recently [8],[10] the electrostatic attraction was balanced exactly by confinement repulsion, which reflects the increase in the kinetic energy of a confined electron. Possible magnetic forces were investigated using the singlet hyperfine ground state of hydrogen [9]. Since the net angular momentum vanishes in that case, the direct magnetic dipole interaction is absent. Instead, the charge is redistributed isotropically and leads to an additional electrostatic force density. Here we add the force generated by the exchange interaction, which has pure quantum character. By generating the exchange hole it induces a full unit charge which exceeds the charges induced by electrostatic polarization.

Notice that only static forces are considered. Dynamical forces, such as oscillations about an equilibrium configuration, are ignored. Although dynamical effects contribute to the energy density [2], they affect the time-averaged force density only via nonlinear (anharmonic) response. That may come into play at distances well below the reduced Compton wavelength as a logarithmic effect, such as the electron self-energy [2].

In Section 2 we start out with a brief introduction to the concept of an exchange hole in solid state physics, followed by its generalization from the Fermi sea to the Dirac sea. An approximate definition of the exchange hole and electron is given which allows analytic work. Section 3 provides the force densities that are generated by the charge densities of the exchange hole, the exchange electron, and the reference electron. This represents a heuristic attempt to capture the effect of the exchange interaction onto an electron in the Dirac sea. It should be taken in the same spirit as the various approximations for the exchange-correlation energy in solids that have been used quite successfully in density functional calculations [3]-[7].

Section 4 makes an attempt at a more rigorous definition of exchange hole and exchange electron using the relativistic three-fermion correlation. It is calculated for both the Fermi and Dirac seas from the Slater determinant of three fermions. A brief summary in Section 5 provides a list of the key findings and makes suggestions for getting closer to establishing a force balance for the electron.

## 2. The Exchange Interaction in the Dirac and Fermi Seas

### 2.1 The Concept of an Exchange Hole

The exchange hole describes the electron density induced by pairwise electron-electron interactions in an electron gas, such as the Fermi sea of a metal. It provides both an intuitive picture and quantitative results about the exchange energy [3]-[7]. One starts with the pair distribution which describes the probability of finding an electron at the point $\mathbf{r}'$ if there is one at $\mathbf{r}$. After selecting the correlations induced by the exchange interaction, this dimensionless probability is multiplied with the electron density to define the exchange hole. It is a correlated one-electron density with dimension (volume)$^{-1}$. The exchange interaction itself originates from the antisymmetry of the fermion pair wave function under the exchange of identical particles. It leads to the exclusion principle which keeps electrons with equal spins apart. Each electron in the Fermi sea is surrounded by an exchange hole with opposite charge and spin.

In quantum electrodynamics the Fermi sea gets replaced by the Dirac sea of electrons and positrons with negative energy that populate the vacuum. The two infinite charge densities exactly compensate each other to keep the vacuum neutral. To generalize the definition of the exchange hole one can take advantage of the practical experience acquired in condensed matter physics about the role of the exchange hole in the Fermi sea. However, significant changes are required for transferring the concept to the Dirac sea. Lorentz invariance needs to be taken into account by combining charge and current density into a four-vector and by replacing the Schrödinger equation with the Dirac equation. One also has to satisfy particle-antiparticle symmetry and conserve charge.

### 2.2 The Exchange Hole in the Fermi Sea

The nonrelativistic definition of the exchange hole starts with the total electron density n($\mathbf{r}$), the key quantity in density functional theory [3]-[7]. It is the sum of the individual probability densities obtained from the one-electron wave functions $\psi_i(\mathbf{r})$:

(1)     n($\mathbf{r}$) = $\Sigma_i^{occ}$ $\psi_i(\mathbf{r})^*\psi_i(\mathbf{r})$      summed over i=1,...,N

The sum is taken over all occupied one-electron states i. In a plane-wave basis the momentum $\mathbf{p}$ and the spin s take the role of the index i. The sum $\Sigma_i$ becomes an integral over $\mathbf{p}$ combined with the sum over the two spin states with $m_s = \pm½ = \uparrow,\downarrow$.

The correlations between two electrons induced by the exchange interaction are extracted from their pair density n($\mathbf{r},\mathbf{r}'$) which is obtained from the antisymmetrized pair wave functions $\psi_{ij}(\mathbf{r},\mathbf{r}')$ and the corresponding pair densities $\psi_{ij}^*\psi_{ij}$. In the following we will consider only electrons with parallel spins which have an antisymmetric spatial wave function. Electrons with opposite spin and a symmetric spatial wave function do not give rise to an exchange term. They are not affected by the exclusion principle.

(2) $\psi_{ij}(\mathbf{r},\mathbf{r}') = [\psi_i(\mathbf{r})\,\psi_j(\mathbf{r}') - \psi_j(\mathbf{r})\,\psi_i(\mathbf{r}')] / \sqrt{2}$

(3) $n(\mathbf{r},\mathbf{r}') = \Sigma_{i\geq j}\,\psi_{ij}(\mathbf{r},\mathbf{r}')^*\,\psi_{ij}(\mathbf{r},\mathbf{r}')$

$= \underbrace{\Sigma_{i\geq j}\,[\psi_i(\mathbf{r})^*\psi_i(\mathbf{r})]\cdot[\psi_j(\mathbf{r}')^*\psi_j(\mathbf{r}')]}_{\tfrac{1}{2}n(\mathbf{r})\cdot n(\mathbf{r}')\ \text{Coulomb}} - \underbrace{\Sigma_{i\geq j}\,[\psi_i(\mathbf{r})^*\psi_j(\mathbf{r})]\cdot[\psi_j(\mathbf{r}')^*\psi_i(\mathbf{r}')]}_{G(\mathbf{r},\mathbf{r}')\ \text{Exchange}}$

The sum over i,j is taken over all unique combinations of occupied one-electron wave functions $\psi_i(\mathbf{r}), \psi_j(\mathbf{r}')$. Double-counting of i,j and j,i is avoided using the sum $\Sigma_{i\geq j}$. It is equal to $\tfrac{1}{2}\Sigma_{i,j}$, since the self-Coulomb term i=j is canceled by the self-exchange term. The first term (Coulomb or Hartree) is a product of one-electron densities, while the second term (exchange or Fock) represents the exchange interaction. Subtracting the Coulomb term isolates the exchange and ensures that $G(\mathbf{r},\mathbf{r}')$ approaches zero for $|\mathbf{r}-\mathbf{r}'|\to\infty$, where the exclusion principle becomes irrelevant.

(4) $G(\mathbf{r},\mathbf{r}') = -\Sigma_{i\geq j}[\psi_i(\mathbf{r})^*\psi_j(\mathbf{r})]\cdot[\psi_j(\mathbf{r}')^*\psi_i(\mathbf{r}')] = -\tfrac{1}{2}\Sigma_{i,j}[\psi_i(\mathbf{r})^*\psi_j(\mathbf{r})]\cdot[\psi_j(\mathbf{r}')^*\psi_i(\mathbf{r}')]$

The minus sign originates from the minus in the antisymmetric pair wave function. Normalization of the correlated pair density $G(\mathbf{r},\mathbf{r}')$ to the uncorrelated Coulomb term yields the dimensionless pair correlation function $g_x$ (often labeled as g–1):

(5) $g_x(\mathbf{r},\mathbf{r}') = \dfrac{\text{Exchange}}{\text{Coulomb}} = -\dfrac{\Sigma_{i\geq j}\,[\psi_i(\mathbf{r})^*\psi_j(\mathbf{r})]\cdot[\psi_j(\mathbf{r}')^*\psi_i(\mathbf{r}')]}{\Sigma_{i\geq j}\,[\psi_i(\mathbf{r})^*\psi_i(\mathbf{r})]\cdot[\psi_j(\mathbf{r}')^*\psi_j(\mathbf{r}')]}$

$g_x(\mathbf{r},\mathbf{r}')$ describes the probability of finding an exchange electron at $\mathbf{r}'$, if there is a reference electron at $\mathbf{r}$. The electron density induced at $\mathbf{r}'$ is given by the product of $g_x(\mathbf{r},\mathbf{r}')$ with the one-electron density $n(\mathbf{r}')$. That defines the exchange hole $n_x(\mathbf{r},\mathbf{r}')$:

(6) $n_x(\mathbf{r},\mathbf{r}') = g_x(\mathbf{r},\mathbf{r}')\cdot n(\mathbf{r}')$

The exchange hole $n_x$ is a one-electron density while the exchange term $G$ is a two-electron density. In a homogeneous electron gas $n_x(\mathbf{r},\mathbf{r}')$ does not depend on $\mathbf{r}$. In that case it takes the form $n_x(\mathbf{r}-\mathbf{r}')$. The following sum rule states that the exchange hole displaces exactly one electron. It can be derived from (6) via the orthonormality of the wave functions $\psi_i(\mathbf{r})$:

(7) $\int n_x(\mathbf{r},\mathbf{r}')\,d^3\mathbf{r}' = -1$    for all $\mathbf{r}$

$\int n_x(\mathbf{r},\mathbf{r}')\,d^3\mathbf{r}' = -\Sigma_{i\geq j}[\psi_i(\mathbf{r})^*\psi_j(\mathbf{r})]\cdot\int\overbrace{[\psi_j(\mathbf{r}')^*\psi_i(\mathbf{r}')]}^{\delta_{ij}}\,d^3\mathbf{r}' / \Sigma_k[\psi_k(\mathbf{r})^*\psi_k(\mathbf{r})] = -1$

The positive charge of the exchange hole neutralizes the charge of an electron in metals over a few atomic distances and thereby greatly shields the effective Coulomb interaction between electrons. That enables the weakly-interacting Fermi liquid picture to survive in metals despite their high electron density.

    A feature that makes the exchange hole particularly attractive for representing electron-electron interactions is the expression for the total exchange energy $U^x$. It is equal to the Coulomb energy between the total charge density $-e\cdot n(\mathbf{r})$ and the charge density of the exchange hole $-e\cdot n_x(\mathbf{r},\mathbf{r}')$:

(8) $\quad U^x = \frac{1}{2} e^2 \cdot \iint n(\mathbf{r}) \frac{1}{|\mathbf{r}-\mathbf{r}'|} n_x(\mathbf{r},\mathbf{r}') \, d^3\mathbf{r}' d^3\mathbf{r}$

$\qquad\quad = -\frac{1}{2} e^2 \cdot \iint \Sigma_{i \geq j} [\psi_i(\mathbf{r})^* \psi_j(\mathbf{r})] \frac{1}{|\mathbf{r}-\mathbf{r}'|} [\psi_j(\mathbf{r}')^* \psi_i(\mathbf{r}')] \, d^3\mathbf{r}' d^3\mathbf{r}$

This is the Hartree-Fock expression for the total exchange energy. The factor ½ avoids double-counting of the same pair of charges in the integrals, once at $\mathbf{r},\mathbf{r}'$ and again at $\mathbf{r}',\mathbf{r}$. Combining (8) with the sum rule (7) shows that the exchange energy tends to be negative (attractive), since $n(\mathbf{r})$ is positive and the integral over $n_x(\mathbf{r},\mathbf{r}')$ negative. One can also define the exchange energy density $u^x$ by omitting the integration over $d^3\mathbf{r}$:

(9) $\quad u^x = \;\; e^2 \cdot \int n(\mathbf{r}) \frac{1}{|\mathbf{r}-\mathbf{r}'|} n_x(\mathbf{r},\mathbf{r}') \, d^3\mathbf{r}'$

$\qquad\quad = -e^2 \cdot \int \Sigma_{i \geq j} [\psi_i(\mathbf{r})^* \psi_j(\mathbf{r})] \frac{1}{|\mathbf{r}-\mathbf{r}'|} [\psi_j(\mathbf{r}')^* \psi_i(\mathbf{r}')] \, d^3\mathbf{r}'$

The exchange hole $n_x$ can be refined by adding other correlations, such as screening. The result is the exchange-correlation hole $n_{xc} = n_x + n_c$. The sum rule (7) remains valid, leaving the correlation hole $n_c$ with no net charge. Screening is typically weaker than exchange [7]. This is particularly noticeable in quantum electrodynamics, where screening corresponds to vacuum polarization (see Section 3). Its electron density is of $O(\alpha^1)$, while that of the exchange hole is of $O(\alpha^0)$.

## 2.3 Generalization to Quantum Electrodynamics

Several changes are needed to transfer the concept of the exchange hole from the non-relativistic Fermi sea to the relativistic Dirac sea. The Schrödinger wave functions $\psi_i$ now become four-component Dirac spinors. Since the Dirac sea represents the vacuum of quantum electrodynamics, its total charge density vanishes. The infinite but opposite charge densities of the electrons and positrons with negative energy cancel each other by symmetry. As a result, the total electron density $n(\mathbf{r})$ defined in (1) vanishes. That would cause a division by zero in (5) and produce trivial results. To solve this problem we focus on the exchange interaction of a specific reference electron (or positron) $\psi_0$ with the vacuum states $\psi_i$. The sum $\Sigma_i^{occ}$ over one-electron densities in (1) is replaced by the electron density of the individual reference electron $\psi_0$. If a positron is chosen as reference, it is counted with a negative sign $s_e$:

(10) $\quad n(\mathbf{r}) \rightarrow n_0(\mathbf{r}) = s_e \cdot \psi_0(\mathbf{r})^* \psi_0(\mathbf{r}) \qquad\qquad s_e = +1\;(-1)\;\; \text{for } e^-\,(e^+)$

This substitution rule produces results similar to Weisskopf's subtraction method, as long as the same reference electron is chosen (for details see the comment with Ref. [2] and the discussion after the definition (17a) of the exchange hole). To maintain relativistic invariance [6], the electron density (10) has to become the time-like component of a relativistic four-current density $n^\mu(\mathbf{r})$. One can also define relativistic scalar and tensor densities $m(\mathbf{r})$ and $s^{\mu\nu}(\mathbf{r})$ which correspond to mass and dipole densities:

(11a) $\quad n^\mu(\mathbf{r}) \rightarrow n_0^\mu(\mathbf{r}) = s_e \cdot \bar{\psi}_0(\mathbf{r}) \gamma^\mu \psi_0(\mathbf{r})$

(11b) $\quad m(\mathbf{r}) \rightarrow m_0(\mathbf{r}) = s_e \cdot \bar{\psi}_0(\mathbf{r}) \psi_0(\mathbf{r})$

(11c) $\quad s^{\mu\nu}(\mathbf{r}) \rightarrow s_0^{\mu\nu}(\mathbf{r}) = s_e \cdot \bar{\psi}_0(\mathbf{r}) \sigma^{\mu\nu} \psi_0(\mathbf{r}) \qquad\quad \sigma^{\mu\nu} = \frac{1}{2} i (\gamma^\mu \gamma^\nu - \gamma^\nu \gamma^\mu)$

The antisymmetric pair wave function contains the outer product of two spinors $\psi_\alpha \otimes \psi_\beta$:

(12) $\quad \psi_{ij}(\mathbf{r},\mathbf{r}') = [\psi_i(\mathbf{r}) \otimes \psi_j(\mathbf{r}') - \psi_j(\mathbf{r}) \otimes \psi_i(\mathbf{r}')]/\sqrt{2}$

The pair density in (3) becomes a four-tensor $\tilde{n}^{\mu\nu}(\mathbf{r},\mathbf{r}')$ consisting of the outer product of two four-vectors. The $\gamma^\mu$ connect wave functions at $\mathbf{r}$, and the $\gamma^\nu$ those at $\mathbf{r}'$. As in (3), the multiplication of $\psi_{ij}$ with its complex conjugate produces Coulomb terms of the form $+(ii \cdot jj)$ and exchange terms of the form $-(ij \cdot ji)$:

(13) $\quad \tilde{n}^{\mu\nu}(\mathbf{r},\mathbf{r}') = \tilde{\Sigma}_{i \geq j} \, \overline{\psi}_{ij}(\mathbf{r},\mathbf{r}') \gamma^\mu \gamma^\nu \psi_{ij}(\mathbf{r},\mathbf{r}') \qquad\qquad \tilde{\Sigma}_{i,j} = \Sigma_{i,j} \, s_{e,i} \cdot s_{e,j}$

$\quad = \underbrace{\tilde{\Sigma}_{i \geq j} [\overline{\psi}_i(\mathbf{r}) \gamma^\mu \psi_i(\mathbf{r})] \cdot [\overline{\psi}_j(\mathbf{r}') \gamma^\nu \psi_j(\mathbf{r}')]}_{\to\; n_0^\mu(\mathbf{r}) \cdot n_0^\nu(\mathbf{r}') \quad \text{Coulomb}} - \underbrace{\tilde{\Sigma}_{i \geq j} [\overline{\psi}_i(\mathbf{r}) \gamma^\mu \psi_j(\mathbf{r})] \cdot [\overline{\psi}_j(\mathbf{r}') \gamma^\nu \psi_i(\mathbf{r}')]}_{\tilde{G}^{\mu\nu}(\mathbf{r},\mathbf{r}') \quad \text{Exchange}}$

The tilde sum contains the sign factors $s_i$ for electrons/positrons from (10). It is carried out over the Dirac sea of electrons and positrons with negative energy (see Figure 1 below). Each of the two vanishing sums in the Coulomb term is replaced by the density $n_0^\mu(\mathbf{r})$ of the reference electron given in (10). That also eliminates double-counting. To isolate the exchange term, the Coulomb term is subtracted from the pair density:

(14a) $\quad \tilde{G}^{\mu\nu}(\mathbf{r},\mathbf{r}') = -\tilde{\Sigma}_{i \geq j}[\overline{\psi}_i(\mathbf{r})\gamma^\mu \psi_j(\mathbf{r})] \cdot [\overline{\psi}_j(\mathbf{r}')\gamma^\nu \psi_i(\mathbf{r}')] = -\tfrac{1}{2}\, \tilde{\Sigma}_{i,j}[\overline{\psi}_i(\mathbf{r})\gamma^\mu \psi_j(\mathbf{r})] \cdot [\overline{\psi}_j(\mathbf{r}')\gamma^\nu \psi_i(\mathbf{r}')]$

(14b) $\quad \tilde{H}^{\mu\nu}(\mathbf{r},\mathbf{r}') = \tfrac{1}{2}\,[\tilde{G}^{\mu\nu}(\mathbf{r},\mathbf{r}') + \tilde{G}^{\nu\mu}(\mathbf{r},\mathbf{r}')] \qquad\qquad\qquad\qquad \text{either } i=0 \text{ or } j=0$

The tensor $\tilde{G}^{\mu\nu}(\mathbf{r},\mathbf{r}')$ needs to be symmetrized, because it contains antisymmetric off-diagonal elements which would lead to unphysical correlations between different components of the four-current. By analogy to (10), one of the two indices $i,j$ is assigned to the reference electron $\psi_0$. This converts the sum $\tilde{\Sigma}_{i,j}$ into two single sums $(\tilde{\Sigma}_{0,j} + \tilde{\Sigma}_{i,0})$.

To define a dimensionless correlation function $\tilde{g}_x(\mathbf{r},\mathbf{r}')$ analogous to (5) would involve a division by the Coulomb term. But the relativistic two-electron densities in (13) are tensors which cannot be divided in Lorentz-invariant fashion. Therefore, the Coulomb denominator is moved to the left side of the equation:

(15) $\quad n_0^\mu(\mathbf{r}) \cdot n_0^\nu(\mathbf{r}') \cdot \tilde{g}_x(\mathbf{r},\mathbf{r}') = \tilde{H}^{\mu\nu}(\mathbf{r},\mathbf{r}')$

The exchange hole $\tilde{n}_x^\mu(\mathbf{r},\mathbf{r}')$ is obtained via (6) by multiplying the dimensionless probability $\tilde{g}_x(\mathbf{r},\mathbf{r}')$ with the relativistic electron density from (10),(11a):

(16) $\quad \tilde{n}_x^\mu(\mathbf{r},\mathbf{r}') = \tilde{g}_x(\mathbf{r},\mathbf{r}') \cdot n_0^\mu(\mathbf{r}') \qquad\qquad \tilde{n}_{x,\nu}(\mathbf{r},\mathbf{r}') = \tilde{g}_x(\mathbf{r},\mathbf{r}') \cdot n_{0,\nu}(\mathbf{r}')$

Multiplication of (15) with $n_{0,\nu}(\mathbf{r}')$ from the right provides an implicit expression for the exchange hole as the time-like component of the four-vector $\tilde{n}_x^\nu(\mathbf{r},\mathbf{r}')$:

(17a) $\quad n_0^\mu(\mathbf{r}) \cdot n_0^\nu(\mathbf{r}') \cdot \underbrace{\tilde{g}_x(\mathbf{r},\mathbf{r}') \cdot n_{0,\nu}(\mathbf{r}')}_{\tilde{n}_{x,\nu}(\mathbf{r},\mathbf{r}')} = \tilde{H}^{\mu\nu}(\mathbf{r},\mathbf{r}') \cdot n_{0,\nu}(\mathbf{r}')$

For the reference electron with $\mathbf{p}=\mathbf{0}$ the spatial components of the four-current $n_0^\mu$ vanish (see (21a)). That leaves a single equation (with $\mu=\nu=0$) defining the exchange hole. This justifies the use of the scalar $\tilde{g}_x(\mathbf{r},\mathbf{r}')$ in (15), rather than a $4^{th}$ rank tensor.

To appreciate the non-trivial sign patterns consider a reference electron in the electron Dirac sea versus a reference positron in the positron Dirac sea. In both cases the tilde sum comes out positive, since the overlap is largest when the reference electron/positron is part of its own Dirac sea. The densities $n_0^\mu(\mathbf{r}), n_0^\nu(\mathbf{r}'), n_{0,\nu}(\mathbf{r}')$ in (17a) have opposite signs for electrons and positrons, which gives the corresponding exchange holes $\tilde{n}_{x,\nu}$ opposite signs, too. For a reference electron with positive energy (as used by Weisskopf [2]) the tilde sum is negative due to its better overlap with the positron Dirac sea. The exchange hole becomes an electron in that case.

Exchange densities with scalar and tensor character can be defined in similar fashion by combining the one-electron densities $m_0(\mathbf{r}), s_0^{\mu\nu}(\mathbf{r})$ from (11b,c) with exchange terms analogous to (14), shown here in abbreviated form:

(17b) $\quad m_0(\mathbf{r}) m_0(\mathbf{r}') \cdot \tilde{m}_x(\mathbf{r},\mathbf{r}') = -\tfrac{1}{2} \tilde{\Sigma}_{i,j} [\overline{\psi}_i(\mathbf{r}) \psi_j(\mathbf{r})] \cdot [\overline{\psi}_j(\mathbf{r}') \psi_i(\mathbf{r}')] \cdot m_0(\mathbf{r}')$

(17c) $\quad s_0^{\mu\nu}(\mathbf{r}) s_0^{\rho\sigma}(\mathbf{r}') \cdot \tilde{s}_{x,\rho\sigma}(\mathbf{r},\mathbf{r}') = -\tfrac{1}{2} \tilde{\Sigma}_{i,j} [\overline{\psi}_i(\mathbf{r}) \sigma^{\mu\nu} \psi_j(\mathbf{r})] \cdot [\overline{\psi}_j(\mathbf{r}') \sigma^{\rho\sigma} \psi_i(\mathbf{r}')] \cdot s_{0,\rho\sigma}(\mathbf{r}')$

Explicit expressions for $n_0^\mu, m_0, s_0^{\mu\nu}$ in (21a,b,c) show that for a wave function $\psi_0$ with $\mathbf{p}=0$ only the components $n_0^0, m_0$, and $s_0^{12} = -s_0^{21}$ survive. That makes the implicit definitions of the corresponding exchange holes in (17a,b,c) explicit. In particular, the result (B7a) for the exchange term on the right side of (17a) becomes proportional to the metric tensor $g^{\mu\nu}$ and thereby reduces both sides to the component $\mu=\nu=0$. The wave function $\psi_0$ of the reference electron can be transformed from its rest frame ($\mathbf{p}=0$) to arbitrary $\mathbf{p}$ values, since the formalism developed here is Lorentz-invariant (including the vacuum state itself). The generalization of the sum rule (7) to quantum electrodynamics will be confirmed by an explicit calculation in (B7a),(23). The exchange interaction of an electron with the Dirac sea displaces indeed exactly one electron: $\int \tilde{n}_x^0(\mathbf{r},\mathbf{r}') d^3\mathbf{r}' = -1$

The expression (8) for the exchange energy can be generalized to:

(18) $\quad U^x = \tfrac{1}{2} e^2 \cdot \iint n^0(\mathbf{r}) \frac{1}{|\mathbf{r}-\mathbf{r}'|} \tilde{n}_x^0(\mathbf{r},\mathbf{r}') d^3\mathbf{r}' d^3\mathbf{r}$

$\qquad = -\tfrac{1}{2} e^2 \cdot \iint \tilde{\Sigma}_{i\geq j} [\overline{\psi}_i(\mathbf{r}) \gamma^0 \psi_j(\mathbf{r})] \frac{1}{|\mathbf{r}-\mathbf{r}'|} [\overline{\psi}_j(\mathbf{r}') \gamma^0 \psi_i(\mathbf{r}')] d^3\mathbf{r}' d^3\mathbf{r}$

The energy density $u^x$ is obtained from by omitting the integration over $\mathbf{r}$:

(19) $\quad u^x = e^2 \cdot \int n^0(\mathbf{r}) \frac{1}{|\mathbf{r}-\mathbf{r}'|} \tilde{n}_x^0(\mathbf{r},\mathbf{r}') d^3\mathbf{r}'$

$\qquad = -e^2 \cdot \int \tilde{\Sigma}_{i\geq j} [\overline{\psi}_i(\mathbf{r}) \gamma^0 \psi_j(\mathbf{r})] \frac{1}{|\mathbf{r}-\mathbf{r}'|} [\overline{\psi}_j(\mathbf{r}') \gamma^0 \psi_i(\mathbf{r}')] d^3\mathbf{r}'$

## 2.4 Calculation of the Exchange Hole

A basis set of plane waves $\psi(\mathbf{r},t) = \psi(\mathbf{r}) \cdot \exp[\pm i p^0 t]$ is used for electrons and positrons with positive and negative energy $p^0$ and spin $s=\uparrow,\downarrow$ (see Appendix B):

(20a) $\quad \psi^e(\mathbf{r},t) = (2\pi \lambda_C)^{-3/2} \cdot u(\mathbf{p},s) \cdot \exp[+i(\mathbf{p}\mathbf{r} - p^0 t)] \qquad p^0 > 0 \qquad$ electrons

$\qquad \psi^e(\mathbf{r},t) = (2\pi \lambda_C)^{-3/2} \cdot \hat{u}(\mathbf{p},s) \cdot \exp[-i(\mathbf{p}\mathbf{r} - p^0 t)] \qquad p^0 < 0 \qquad$ electrons

$\qquad \psi^p(\mathbf{r},t) = (2\pi \lambda_C)^{-3/2} \cdot v(\mathbf{p},s) \cdot \exp[-i(\mathbf{p}\mathbf{r} - p^0 t)] \qquad p^0 > 0 \qquad$ positrons

$\qquad \psi^p(\mathbf{r},t) = (2\pi \lambda_C)^{-3/2} \cdot \hat{v}(\mathbf{p},s) \cdot \exp[+i(\mathbf{p}\mathbf{r} - p^0 t)] \qquad p^0 < 0 \qquad$ positrons

(20b) $\langle i|j\rangle = \Sigma_\alpha \int d^3\mathbf{r}\ \psi(\mathbf{r})^*_{\mathbf{p},p^0,s,\alpha}\cdot\psi(\mathbf{r})_{\mathbf{p}',p^{0'},s',\alpha}$

$\quad\quad\quad = \delta_{s,s'}\cdot\delta_{\text{sign}(p^0),\text{sign}(p^{0'})}\cdot\delta^3(\mathbf{p}/m_e - \mathbf{p}'/m_e)$ orthonormality

(20c) $\tilde\Sigma_i |i\rangle\langle i| = \text{sign}(p^0)\cdot\Sigma_s \int d^3(\mathbf{p}/m_e)\ [\psi^e(\mathbf{r})_{\mathbf{p},p^0,s,\alpha}\cdot\bar\psi^e(\mathbf{r}')_{\mathbf{p},p^0,s,\beta} - \psi^p(\mathbf{r})_{\mathbf{p},p^0,s,\alpha}\cdot\bar\psi^p(\mathbf{r}')_{\mathbf{p},p^0,s,\beta}]$

$\quad\quad\quad = \delta_{\alpha\beta}\cdot\delta^3(\mathbf{r}-\mathbf{r}')\cdot(m_e/E)$ completeness

(20d) $\psi_0(\mathbf{r}) = (2\pi\lambda_C)^{-3/2}\cdot\hat{u}(\mathbf{0},\uparrow)\quad \hat{u}(\mathbf{0},\uparrow) = \{0,0,1,0\}\quad p^0 = -m_e$ reference electron

The summation index i represents $\mathbf{p},s$ and $e^-,e^+$. Spinor components are labeled $\alpha,\beta$. The sum over $\mathbf{p}$ is converted to the dimensionless integral $\int d^3(\mathbf{p}/m_e)$. The relativistic summation factor $(2E)^{-\frac{1}{2}}$ has been incorporated into the Dirac spinors $u(\mathbf{p},s),\hat{u}(\mathbf{p},s)$ and $v(\mathbf{p},s),\hat{v}(\mathbf{p},s)$ which are defined in Appendix B. The covariant sum over the mass shell has the form $\tilde\Sigma_i|i\rangle\langle i| = \int d^4(p/m_e)\ \delta(p^\mu p_\mu - m_e^2)\cdot\text{sign}(p^0)\cdot|i\rangle\langle i| = (m_e/2E)\cdot\int d^3(\mathbf{p}/m_e)\cdot|i\rangle\langle i|$ (see [12]). This can be seen from the relation $\delta(p^\mu p_\mu - m_e^2) = \delta[(p^0+E)\cdot(p^0-E)] = \delta(p^0-E)/2E$ using $E=(m_e^2+\mathbf{p}^2)^{\frac{1}{2}}$. The reference electron in the Dirac sea can be chosen to be in its rest frame ($\mathbf{p}=\mathbf{0},p^0=-m_e$). The choices $s=\uparrow,\downarrow$ are equivalent, because the Dirac sea is unpolarized.

The one-electron densities (11a,b,c) become normalization constants, with their signs depending on whether one chooses an electron or positron as reference:

(21a) $n_0^\mu(\mathbf{r}) = s_e\cdot\bar\psi_0\gamma^\mu\psi_0\ = +s_e\cdot(2\pi\lambda_C)^{-3}\cdot\delta^{\mu 0}\quad\quad\quad s_e=1$ for $e^-$ $(-1$ for $e^+)$

(21b) $m_0(\mathbf{r}) = s_e\cdot\bar\psi_0\psi_0\ = -s_e\cdot(2\pi\lambda_C)^{-3}\cdot 1$

(21c) $s_0^{\mu\nu}(\mathbf{r}) = s_e\cdot\bar\psi_0\sigma^{\mu\nu}\psi_0 = -s_e\cdot(2\pi\lambda_C)^{-3}\cdot(\delta^{\mu 1}\delta^{\nu 2} - \delta^{\mu 2}\delta^{\nu 1}) = -s_e\cdot 2\bar\psi_0\sigma^3\psi_0$

The spatial components $n_0^i$ vanish, since $\mathbf{p}=\mathbf{0}$ for $\psi_0$. Likewise, only $s_0^{12} = -s_0^{21}$ survives.

The structure of the biquadratic terms defining the exchange hole in (14),(17a) is illustrated in Figure 1 below. Since one of the indices i,j represents the reference wave function $\psi_0$, the summation is restricted to the crossed lines $p_i^0 = -m_e$ and $p_j^0 = -m_e$:

$\tilde G^{\mu\nu}(\mathbf{r},\mathbf{r}') = -\tfrac{1}{2}\tilde\Sigma_{i,j}[\bar\psi_i(\mathbf{r})\gamma^\mu\psi_j(\mathbf{r})]\cdot[\bar\psi_j(\mathbf{r}')\gamma^\nu\psi_i(\mathbf{r}')]$

$\quad\quad\quad = -\tfrac{1}{2}[\tilde\Sigma_i(i0\cdot 0i) + \tilde\Sigma_j(0j\cdot j0)]$

In (B7a) this expression is decomposed into a product of the spinors $\hat{u}(\mathbf{p},s),\hat{v}(\mathbf{p},s),\hat{u}(\mathbf{0},\uparrow)$ with the plane waves $e^{\pm i(\mathbf{p}\mathbf{r}-p^0 t)}$ and symmetrized according to (14b). One still has to include the normalization factors from (20a,d) and (21a) which contribute a factor of $(2\pi\lambda_C)^{-3} = m_e^3(2\pi)^{-3}$ for each pair of spinors $\psi^*\psi$. The summation $\Sigma_i = \int d^3(\mathbf{p}/m_e)$ becomes an inverse cosine Fourier transform which is carried out in (D5). After adding the vanishing sine transform the result takes the standard form (A3):

(22) $\tilde n_x^h(\mathbf{r},\mathbf{r}') = \tilde n_x^0(\mathbf{r},\mathbf{r}') = -\lambda_C^{-3}\cdot(2\pi)^{-3}\int\dfrac{m_e}{E(\mathbf{p})}\cdot e^{i\mathbf{p}(\mathbf{r}-\mathbf{r}')}\ d^3(\mathbf{p}/m_e)\quad\quad E(\mathbf{p}) = (m_e^2+\mathbf{p}^2)^{\frac{1}{2}}$

The evaluation yields the exchange hole $\tilde n_x^h$ in terms of the modified Bessel function $K_1$:

(23a) $\tilde n_x^h(r) = -\lambda_C^{-3}\cdot 1/2\pi^2\cdot z^{-1}K_1(z)\quad\quad\quad z = r/\lambda_C = m_e r\quad\quad r = |\mathbf{r}-\mathbf{r}'|$

(23b) $\quad 4\pi r^2 \cdot \tilde{n}_x^h(r) \to -2/\pi \qquad$ for $r \to 0 \quad (\lambda_C = 1)$

$\qquad 4\pi r^2 \cdot \tilde{n}_x^h(r) \to -(2/\pi)^{1/2} \cdot r^{1/2} e^{-r} \quad$ for $r \to \infty \quad (\lambda_C = 1)$

$\qquad \int_0^\infty 4\pi r^2 \cdot \tilde{n}_x^h(r) \, dr = -1$

(24) $\quad \int_0^\infty r \cdot 4\pi r^2 \cdot \tilde{n}_x^h(r) \, dr = 4/\pi \cdot \lambda_C$

This is the average distance of the exchange hole from the reference electron. The radial electron density of the exchange hole is shown in Figure 2 below.

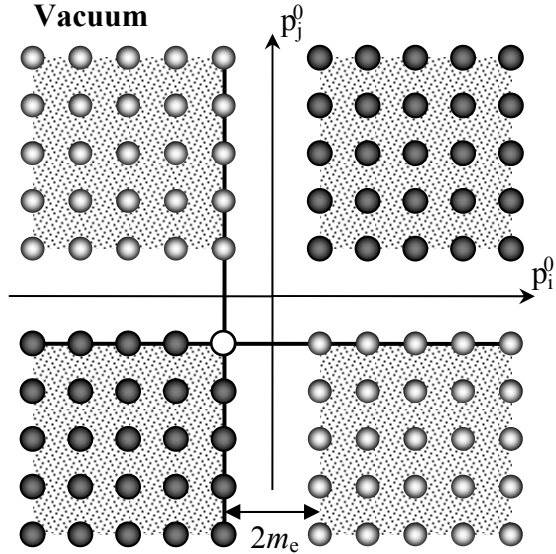

**Figure 1** Schematic visualization of the tilde sum $\tilde{\Sigma}_{i,j}$ in the biquadratic exchange term $\tilde{G}^{\mu\nu}(\mathbf{r},\mathbf{r}')$. Dark dots represent plus signs ($e^-e^-, e^+e^+$) and light dots stand for minus signs ($e^-e^+, e^+e^-$). Box quantization yields a grid of states in the allowed energy regions with $|p^0| \geq m_e$ (shaded). The states $\psi_i, \psi_j$ are shown as dots in the plane spanned by their energies $p_i^0, p_j^0$. The exchange terms $(i0 \cdot 0i), (0j \cdot j0)$ of the reference electron $\psi_0$ are highlighted by crossed lines. Self-exchange and self-Coulomb terms cancel each other (open circle).

## 2.5 The Exchange Electron

Charge conservation requires a negative counter-charge to the exchange hole, such that the response of the vacuum to the reference electron remains neutral. This leads to an exchange electron which describes the response of the Dirac sea to the exchange hole. In addition to the reference electron and the exchange hole one has now a third object. Consequently, the pair correlation needs to be generalized to a three-fermion correlation. Before getting involved with the three-body problem in Section 4 we use three pair correlations as approximation. Each of three pairings acquires an exchange hole and an exchange electron. Furthermore, the exchange electron is approximated by successive pair correlations, one converting the reference electron into the exchange hole and the other converting the exchange hole into the exchange electron. The resulting self-convolution of the exchange hole defines the exchange electron. Its Fourier transform is the square of the Fourier transform of the exchange hole in (22):

(25) $\quad \tilde{n}_x^e(\mathbf{r},\mathbf{r}') = +\lambda_C^{-3} \cdot (2\pi)^{-3} \int \left[\dfrac{m_e}{E(\mathbf{p})}\right]^2 \cdot e^{i\mathbf{p}(\mathbf{r}-\mathbf{r}')} \, d^3(\mathbf{p}/m_e)$

The inverse Fourier transform is carried out in (D6). The result is rather simple:

(26a) $\quad \tilde{n}_x^e(r) = \lambda_C^{-3} \cdot 1/4\pi \cdot z^{-1} e^{-z} \qquad\qquad z = r/\lambda_C = m_e r \qquad r = |\mathbf{r}-\mathbf{r}'|$

(26b) $4\pi r^2 \cdot \tilde{n}_x^e(r) \to r$      for $r \to 0$    ($\lambdabar_C = 1$)

$4\pi r^2 \cdot \tilde{n}_x^e(r) \to r \cdot e^{-r}$    for $r \to \infty$    ($\lambdabar_C = 1$)

$\int_0^\infty 4\pi r^2 \cdot \tilde{n}_x^e(r)\, dr = +1$

(27) $\int_0^\infty r \cdot 4\pi r^2 \cdot \tilde{n}_x^e(r)\, dr = 2 \cdot \lambdabar_C$

The last line provides the average distance of the exchange electron from the reference electron. Its distance from the hole is given by (24). Exchange electron and exchange hole are compared in Figure 2.

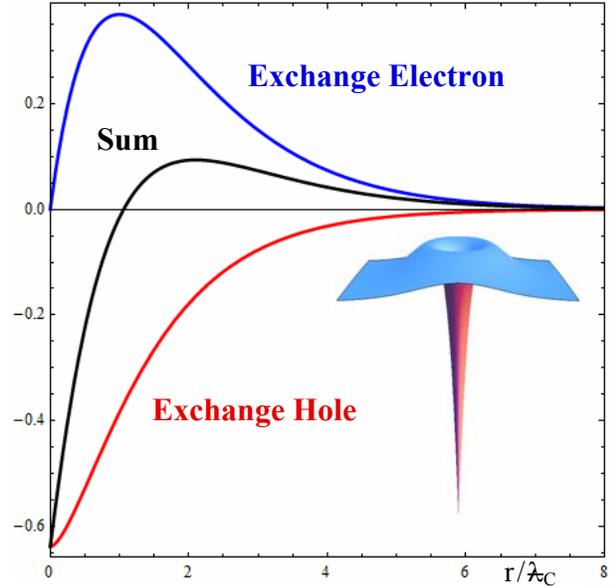

**Figure 2** Radial electron densities of the exchange hole $4\pi r^2 \cdot \tilde{n}_x^h(r)$ in (23), the exchange electron $4\pi r^2 \cdot \tilde{n}_x^e(r)$ in (26), and their sum (also shown as inset). The switch from hole to electron occurs near the reduced Compton wavelength $\lambdabar_C$. The integrals of the densities over $d^3\mathbf{r}$ are $-1, +1,$ and $0$.

The distances between reference electron, exchange hole, and exchange electron obtained from their pair correlations suggest a triangular arrangement similar to that in the negative ion of positronium (see (50),(51) and [14]).

    The convolution that created the exchange electron from the exchange hole can be iterated to include higher order correlations. That creates a series of alternating holes $\tilde{n}_x^{h,n}$ and electrons $\tilde{n}_x^{e,n}$ which can be summed to infinity ($\tilde{n}_x^\infty$), as shown in Figure 3. To ensure charge neutrality, finite sums need to be truncated after an exchange electron. One obtains the following densities, with $n = 1,2,3,...,\infty$:

(28) $\tilde{n}_x^{h,n}(\mathbf{r},\mathbf{r}') = (2\pi)^{-3} \cdot \int [-E(\mathbf{p})]^{-2n+1} \cdot e^{i\mathbf{p}(\mathbf{r}-\mathbf{r}')}\, d^3\mathbf{p}$     $\lambdabar_C^{-1} = m_e = 1$    $E(\mathbf{p}) = (1+\mathbf{p}^2)^{1/2}$

(29) $\tilde{n}_x^{e,n}(\mathbf{r},\mathbf{r}') = (2\pi)^{-3} \cdot \int [-E(\mathbf{p})]^{-2n} \cdot e^{i\mathbf{p}(\mathbf{r}-\mathbf{r}')}\, d^3\mathbf{p}$

(30) $\tilde{n}_x^\infty(\mathbf{r},\mathbf{r}') = (2\pi)^{-3} \cdot \int -[1+E(\mathbf{p})]^{-1} \cdot e^{i\mathbf{p}(\mathbf{r}-\mathbf{r}')}\, d^3\mathbf{p}$

(31) $4\pi r^2 \cdot \tilde{n}_x^{h,n}(r) = -r^n\, K_{n-2}(r) / [2^{n-2}\, \Gamma(\tfrac{1}{2})\, \Gamma(n-\tfrac{1}{2})]$

(32) $4\pi r^2 \cdot \tilde{n}_x^{e,n}(r) = +r^{n+\tfrac{1}{2}}\, K_{n-3/2}(r) / [2^{n-3/2}\, \Gamma(\tfrac{1}{2})\, \Gamma(n)]$

(33) $4\pi r^2 \cdot \tilde{n}_x^\infty(r) \approx -2/\pi \cdot \exp(-4/\pi \cdot r)$

Half-integer Bessel functions for the exchange electrons can be converted to exponential functions. For the inverse Fourier transform of the infinite sum we provide a simple approximation which matches the integral together with the values at $r_1=0$ and $r_2 \approx \lambda_C$. The integral $-½$ is obtained from the value of the Fourier transform at $\mathbf{p}=0$. That is the average between a series truncated after a hole and one truncated after an electron. As shown in Fig. 3, the exchange electron gets pushed out to infinity for $N \to \infty$. This disappearance of half an electron toward $r=\infty$ is reminiscent of the exchange hole in the Fermi sea, where the displaced electron is moved to $r=\infty$ via an empty state just above the Fermi level. One has to keep in mind, though, that such a series of iterated pair correlations may be a poor representation of the true multi-fermion correlation.

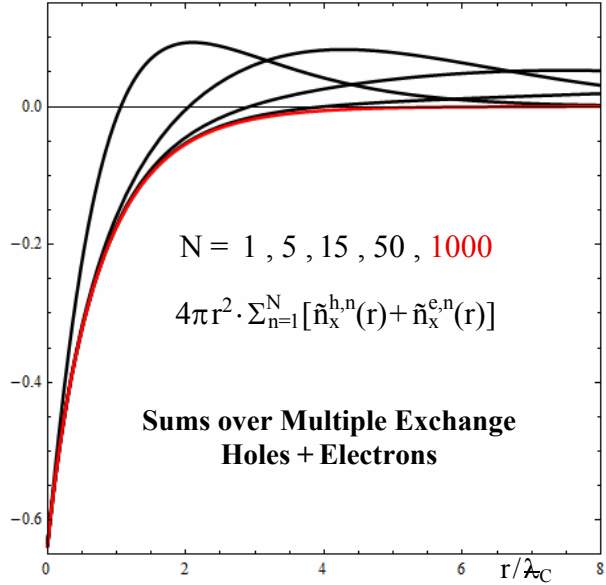

**Figure 3** Higher order correlations, obtained by repeated convolutions of the exchange hole in (31) and (32). The radial electron densities are shown for sums over N exchange holes and electrons (compare Figure 2 for N=1). The negative hole part approaches a finite limit for $N \to \infty$ given by (33). But the positive electron part becomes spread out over an infinitely-large range with an amplitude approaching zero. The integral over the remaining exchange hole approaches $-½$ for $N \to \infty$.

### 2.6 Comparison with the Fermi Sea

Before concluding the discussion of the exchange hole in the Dirac sea, it is appropriate to compare it with the exchange hole in the Fermi sea from condensed matter physics. Figure 4 (below) adds the electrons of the Fermi sea to the Dirac sea. They replace positrons with negative energy, which stood for empty electron states. The electrons of the Fermi sea appear as extra dark dots with positive energies $p_i^0$ or $p_j^0$, extending up to the Fermi level $E_F$. The reference electron for the Fermi sea has the positive energy $+m_e$ with $\mathbf{p}=0$, as indicated by crossed full lines in Fig. 4. Calculations of the exchange hole in the Fermi sea ignore the Dirac sea [4] using the "no pair" approximation – even when they are relativistic [6].

The definitions of the pair correlation $\tilde{g}_x$ in (15) and of the exchange hole $\tilde{n}_x^0$ in (17a) are general enough to be applied to the Fermi sea. The sum over the Dirac sea is replaced by a sum over the electron spinors $u(\mathbf{p},s)$ with $m_e \leq p^0 \leq E_F$. They replace positron spinors $\hat{v}(\mathbf{p},s)$ with $-m_e \geq p^0 \geq -E_F$ and thereby change the sign factor $s_e$. The $\mathbf{p}$-integration in (B7a) is cut off at the Fermi momentum $p_F$. In the limit $E(\mathbf{p}) \approx m_e$, $p_F \ll m_e$ the integrand approaches 1. This adaptation of (17a) to the Fermi sea produces the exchange hole $n_x^F$:

(34)  $n_x^F(r) = -1/6\pi^2 \cdot \{3z^{-3}[\sin(z) - z \cdot \cos(z)]\} \cdot p_F^3$   $z = p_F \cdot |\mathbf{r}|$

$\int_0^R 4\pi r^2 \cdot n_x^F(r) d^3\mathbf{r} \to -1 + 2/\pi \cdot \sin(R)$  for $R \to \infty$   sum rule

The function $n_x^F(r)$ is known as the density matrix in electron gas physics [4],[6]. The actual exchange hole $g^F(r)$ contains the square of $n_x^F(r)$:

(35) $\quad g^F(r) = -1/6\pi^2 \cdot \{3z^{-3}[\sin(z) - z\cdot\cos(z)]\}^2 \cdot p_F^3$

$\quad\quad \int_0^\infty 4\pi r^2 \cdot g^F(r) d^3\mathbf{r} = -1$

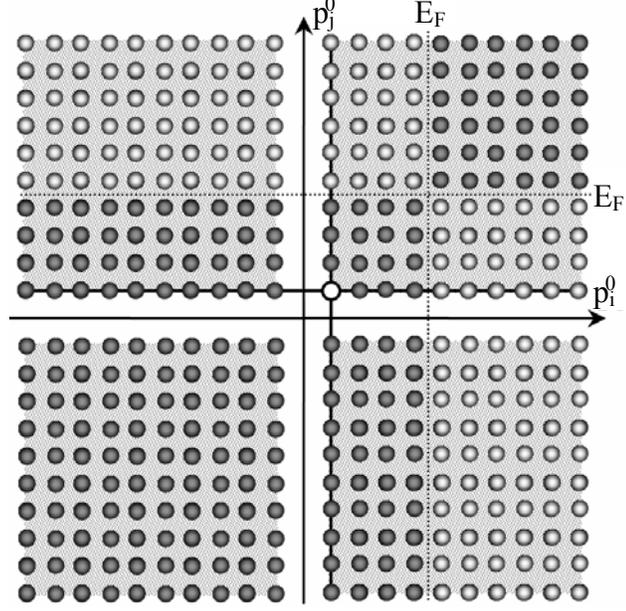

**Figure 4** Similar to Figure 1, but with the Fermi sea added to the Dirac sea. Electrons in the Fermi sea are represented by dark dots with $m_e \leq p^0 \leq E_F$ in the upper right quadrant. They replace positrons with negative energy which are identified with empty electron states. The usual definition of the exchange hole in the Fermi sea ignores the presence of the Dirac sea and therefore lacks the Dirac exchange hole. The crossed lines mark the reference electron in the Fermi sea. That was also used for the Dirac sea by Weisskopf [2].

Figure 5 compares the exchange holes $n_x^F$ and $g^F$ defined in (34) and (35). They are comparable, but not identical. $n_x^F$ is wider, and its sum rule integral exhibits undamped oscillations at large r (see the second line of (34)). These discrepancies originate from different summation methods. $g^F$ is integrated over all pairs **p,p′**, while $n_x^F$ involves only a single **p**-integral. **p′** was set to zero for the Dirac sea to represent the reference electron. It is possible to perform a double integration for the Dirac sea, too, as shown in (B8)-B(10). The result is an exchange hole proportional to $-|\tilde{n}_x^h|^2$, analogous to $g^F$ being proportional to $-|n_x^F|^2$ in the Fermi sea. But $-|n_x^h|^2$ is not normalizable at r=0 due to its $r^{-4}$ divergence. We will return to this topic in Section 4.1 where the exchange hole and electron are both defined via the relativistic three-fermion correlation. Applying that definition to the Fermi sea in Section 4.2 produces the standard exchange hole $g^F$, while $|n_x^F|$ becomes the exchange electron.

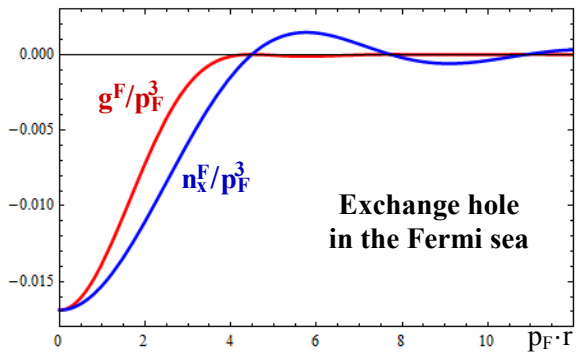

**Figure 5** The standard exchange hole $g^F$ in the Fermi sea from (35), compared to the hole $n_x^F$ obtained in (34) by adapting the definition (17a) for the Dirac sea to the Fermi sea. $g^F$ is proportional to $-|n_x^F|^2$. Both satisfy the sum rule (7), but the integral of $4\pi r^2 \cdot n_x^F(r)$ oscillates (see (34)). $p_F = 2\pi/\lambda_F = (2m_e E_F)^{1/2}$

## 3. Exchange Force Densities

The exchange interactions between the exchange hole (h), the exchange electron (e), and the reference electron (r) create a variety of electrostatic force densities of the form $\rho \cdot \mathbf{E}$. The charge density $\rho(r) = -e \cdot n(r)$ is derived from an electron density $n(r)$, while $\mathbf{E}(r)$ is the electric field derived from $\rho$. Various combinations of $\mathbf{E}^h, \mathbf{E}^e, \mathbf{E}^r$ with $\rho^h, \rho^e$ are shown in Figures 6 and 7 below. They are multiplied by the factor $4\pi r^2$ to reduce their strong $\mathbf{r}$-dependence and thereby distinguish them more clearly. The fields $\mathbf{E}^h, \mathbf{E}^e$ are derived from the electron densities in (23),(26) by solving Poisson's equation and taking the gradients of the resulting potentials $\Phi^h, \Phi^e$:

(36) $\quad \nabla^2 \Phi(r) = r^{-1}[r \cdot \Phi(r)]'' = -4\pi \cdot [-e \cdot n(r)] \qquad n(r) = \tilde{n}_x^h(r), \tilde{n}_x^e(r)$

$\qquad \mathbf{E}(r) = -\nabla \Phi(r)$

(37) $\quad \Phi^h(r) = +e \cdot [K_0(r) \cdot \mathbf{L}_{-1}(r) + K_1(r) \cdot \mathbf{L}_0(r)]$

$\qquad \mathbf{E}^h(r) = +e \cdot r^{-1}[K_0(r) \cdot (\mathbf{L}_{-1}(r) + 2/\pi) + K_1(r) \cdot \mathbf{L}_0(r)] \cdot \mathbf{e}_r$

(38) $\quad \Phi^e(r) = -e \cdot r^{-1}(1 - e^{-r})$

$\qquad \mathbf{E}^e(r) = -e \cdot r^{-2}[(1 - e^{-r}(1+r)] \cdot \mathbf{e}_r$

A prime denotes the derivative with respect to $r$. The $K_n$ and $\mathbf{L}_m$ are modified Bessel and Struve functions (see Appendix D and Ref. [15]).

It is also interesting to compare these exchange force densities with those generated by the Coulomb interaction. The dominant contribution comes from the charge density $\rho^{VP} = -e \cdot n^{VP}$ of the vacuum polarization induced by the reference electron. It is combined with the Coulomb field $\mathbf{E}^r$ of the reference electron, which is localized at the point $\mathbf{r}=\mathbf{0}$ by definition. The charge density and electric field created by vacuum polarization are derived from the potential $\Phi^{VP}$ using results from [10],[11]:

(39) $\quad n^{VP}(r) = -\frac{2\alpha}{3\pi^2}\{Ki_1(z) + z^{-1}K_0(z) - (1 - 2z^{-2}) \cdot K_1(z)]\} \qquad z = 2r$

$\qquad = -\frac{\alpha}{3\pi}\{1 - K_0(z) \cdot [z \cdot \mathbf{L}_{-1}(z) - \frac{2}{\pi}z^{-1}] - K_1(z) \cdot [z \cdot \mathbf{L}_0(z) + \frac{2}{\pi}(1 - 2z^{-2})]\}$

(40) $\quad \mathbf{E}^{VP}(r) = e \cdot \frac{4\alpha}{9\pi} z^{-2}[-z^3 \cdot Ki_1(z) - (z^2 - 6) \cdot K_0(z) + (z^3 + z) \cdot K_1(z)] \cdot \mathbf{e}_r$

$\qquad = e \cdot \frac{2\alpha}{9}\{-z + K_0(z) \cdot [z^2 \cdot \mathbf{L}_{-1}(z) - \frac{2}{\pi}(1 - 6z^{-2})]$

$\qquad \qquad + K_1(z) \cdot [z^2 \cdot \mathbf{L}_0(z) + \frac{2}{\pi}(z + z^{-1})]\} \cdot \mathbf{e}_r$

The electron density $\rho^{VP}$ generated by the Coulomb interaction is of $O(\alpha)$. That is small compared to the electron densities of $O(1)$ created by the exchange interaction. The corresponding force densities of $O(\alpha^2)$ and $O(\alpha)$. This is reminiscent of the exchange hole versus the Coulomb hole (or correlation hole) in condensed matter physics. The latter is significantly weaker, and its total charge vanishes. The strength of the exchange versus Coulomb forces comes out clearly in Figure 7. Since vacuum polarization involves the creation of virtual electron-positron pairs, its decay length at large distances is $\frac{1}{2}\lambdabar_C$ (compared to $\lambdabar_C$ for the exchange hole and electron). Vacuum polarizatiom becomes significant only at distances much smaller than $\lambdabar_C$.

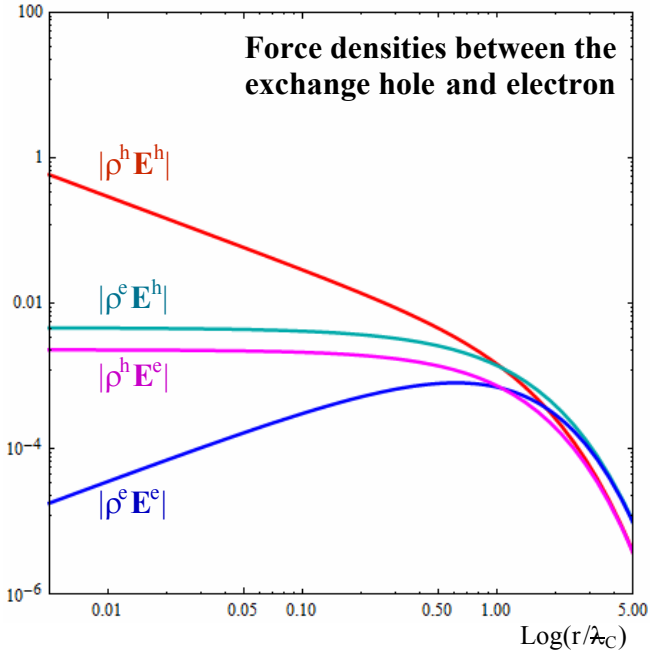

**Figure 6**  Force densities created by the electric fields $\mathbf{E}^h$ and $\mathbf{E}^e$ of the exchange hole and exchange electron, acting on their charge densities $\rho^h$ and $\rho^e$ (multiplied by $4\pi r^2$ for better distinction). There are two cross-wise interactions and two self-interactions. The latter are allowed, because the exchange hole and electron are many-body objects formed by electrons and positrons from the Dirac sea. Above $\lambdabar_C$ these forces compete with those induced by the Coulomb field $\mathbf{E}^r$ of the reference electron in Figure 7, but below $\lambdabar_C$ they become negligible.

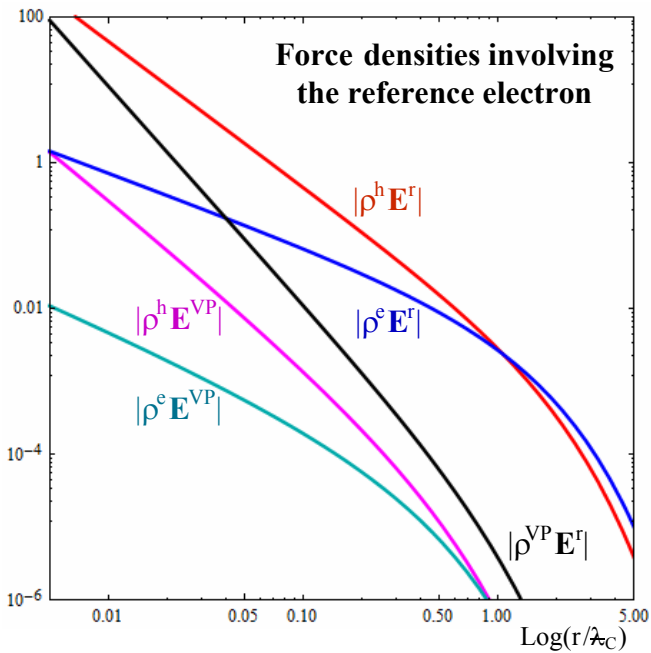

**Figure 7**  Analogous to Figure 6, but for the force densities involving the Coulomb field $\mathbf{E}^r$ of the reference electron and the field $\mathbf{E}^{VP}$ produced by its vacuum polarization. For $r<\lambdabar_C$ the attraction toward the exchange hole exceeds the repulsion from the more distant exchange electron (see the charge densities in Figure 2). The forces related to vacuum polarization are of higher order in $\alpha$ and do not play a significant role near $\lambdabar_C$. That illustrates the dominant role of the exchange interaction of an electron with the Dirac sea (compared to its Coulomb interaction).

For analyzing all these force densities it is helpful to first consider pairwise forces within the exchange exciton. As shown in Figure 6, there are two attractive forces acting between the exchange hole and electron, plus two repulsive self-interactions. The situation becomes more complicated when the reference electron gets involved. Instead of pairwise forces derived from pair correlations one has to deal with correlated three-center forces. For example, the triangular configuration of reference electron, exchange hole, and exchange electron suggested in (50) creates three different charge centers connected by forces along different directions. Such three-body correlations will be topic of the next section.

## 4. An Exchange Exciton from Three-Fermion Correlations

### 4.1 Definition of the Exchange Exciton

While the pair density is sufficient to describe the reference electron and its exchange hole, the three-fermion density is needed to properly include the exchange electron. The response of the Dirac sea to the reference electron now becomes an electron-hole pair, the exchange exciton. In the following, we generalize the definition of the exchange hole from the pair correlation from Section 2.3 to the definition of the exchange exciton from the three-fermion correlation. The antisymmetric wave function for 3 electrons analogous to (12) is a 3×3 Slater determinant involving the outer products of Dirac spinors [6]:

$$(41) \quad \psi_{ijk}(\mathbf{r},\mathbf{r}',\mathbf{r}'') = \begin{vmatrix} \psi_i(\mathbf{r}) & \psi_j(\mathbf{r}) & \psi_k(\mathbf{r}) \\ \psi_i(\mathbf{r}') & \psi_j(\mathbf{r}') & \psi_k(\mathbf{r}') \\ \psi_i(\mathbf{r}'') & \psi_j(\mathbf{r}'') & \psi_k(\mathbf{r}'') \end{vmatrix} / \sqrt{6} = \Sigma_{i,j,k}\, \varepsilon^{ijk} \cdot \psi_i(\mathbf{r}) \otimes \psi_j(\mathbf{r}') \otimes \psi_k(\mathbf{r}'') / \sqrt{6}$$

The relativistic three-fermion charge-current density becomes a third-rank tensor which depends on three wave functions and three spatial coordinates:

$$(42) \quad \tilde{n}^{\mu\nu\lambda}(\mathbf{r},\mathbf{r}',\mathbf{r}'') = \tfrac{1}{6} \tilde{\Sigma}_{i,j,k}\, \overline{\psi}_{ijk}(\mathbf{r},\mathbf{r}',\mathbf{r}'')\gamma^\mu\gamma^\nu\gamma^\lambda \psi_{ijk}(\mathbf{r},\mathbf{r}',\mathbf{r}'') \qquad \tilde{\Sigma}_{i,j,k} = \Sigma_{i,j,k}\, s_{e,i} \cdot s_{e,j} \cdot s_{e,k}$$

The tilde sum $\tilde{\Sigma}_{i,j,k}$ is taken over the Dirac sea of negative energy electrons and positrons, with a minus sign $s_i$ for each pair of positron wave functions. The Dirac matrices $\gamma^\mu, \gamma^\nu, \gamma^\lambda$ connect wave functions at $\mathbf{r},\mathbf{r}',\mathbf{r}''$ respectively. The factor $\tfrac{1}{6}$ accounts for multiple counting of the 6 permutations of the wave functions $i,j,k$ (analogous to the factor $\tfrac{1}{2}$ for double-counting in (14)). The multiplication of $\psi_{ijk}$ with $\overline{\psi}_{ijk}$ produces the three-electron density $\tilde{n}^{\mu\nu\lambda}$ analogous to (13). It consists of Coulomb terms of the form $(ii \cdot jj \cdot kk)$ and exchange terms $\tilde{G}^{\mu\nu\lambda}$ with flipped indices. The latter describe the three-fermion correlations induced by the exchange interaction:

$$(43) \quad \tilde{n}^{\mu\nu\lambda}(\mathbf{r},\mathbf{r}',\mathbf{r}'') = \tfrac{1}{6}\, n^\mu(\mathbf{r}) n^\nu(\mathbf{r}') n^\lambda(\mathbf{r}'') + \tilde{G}^{\mu\nu\lambda}(\mathbf{r},\mathbf{r}',\mathbf{r}'')$$

$$(44) \quad \tilde{G}^{\mu\nu\lambda}(\mathbf{r},\mathbf{r}',\mathbf{r}'') = -\tfrac{1}{6} \tilde{\Sigma}_{i,j,k}\, [\overline{\psi}_i(\mathbf{r})\gamma^\mu \psi_j(\mathbf{r})] \cdot [\overline{\psi}_j(\mathbf{r}')\gamma^\nu \psi_i(\mathbf{r}')] \cdot [\overline{\psi}_k(\mathbf{r}'')\gamma^\lambda \psi_k(\mathbf{r}'')] + \ldots$$
$$= -\tfrac{1}{6} \tilde{\Sigma}_{i,j,k} (i\widehat{j \cdot j}i \cdot kk) + \tfrac{1}{6} \tilde{\Sigma}_{i,j,k} (ij \cdot \widehat{ki} \cdot jk) \quad + \quad \text{permutations} \quad (\text{see B), C) below})$$

A scalar three-fermion correlation function $\tilde{g}_3(\mathbf{r},\mathbf{r}',\mathbf{r}'')$ analogous to the pair correlation $\tilde{g}_x(\mathbf{r},\mathbf{r}')$ in (15) can be defined implicitly by moving the product of one-electron densities to the left side and using the tensor $\tilde{H}^{\mu\nu\lambda}$, the symmetrized version of $\tilde{G}^{\mu\nu\lambda}$ (see (14a,b)):

$$(45) \quad n^\mu(\mathbf{r}) n^\nu(\mathbf{r}') n^\lambda(\mathbf{r}'') \cdot \tilde{g}_3(\mathbf{r},\mathbf{r}',\mathbf{r}'') = \tilde{H}^{\mu\nu\lambda}(\mathbf{r},\mathbf{r}',\mathbf{r}'')$$

$\tilde{g}_3(\mathbf{r},\mathbf{r}',\mathbf{r}'')$ is the dimensionless probability to find fermions at $\mathbf{r}'$ and $\mathbf{r}''$, if there is one at $\mathbf{r}$. $\tilde{G}^{\mu\nu\lambda}(\mathbf{r},\mathbf{r}',\mathbf{r}'')$ has the dimension $(\text{volume})^{-3}$. To preserve the dimension $(\text{volume})^{-1}$ of a one-electron density for the exchange exciton $\tilde{n}_3^\mu$, the three-fermion correlation $\tilde{g}_3$ needs to be multiplied with the one-electron density $n^\mu$:

$$(46) \quad \tilde{n}_3^\mu(\mathbf{r},\mathbf{r}',\mathbf{r}'') = \tilde{g}_3(\mathbf{r},\mathbf{r}',\mathbf{r}'') \cdot n^\mu(\mathbf{r}'') \qquad\qquad \tilde{n}_{3,\lambda}(\mathbf{r},\mathbf{r}',\mathbf{r}'') = \tilde{g}_3(\mathbf{r},\mathbf{r}',\mathbf{r}'') \cdot n_\lambda(\mathbf{r}'')$$

That leads to a contraction of (45) with $n_\lambda(\mathbf{r}'')$ which defines the exchange exciton $\tilde{n}_{3,\lambda}(\mathbf{r},\mathbf{r}',\mathbf{r}'')$ implicitly, as in (16) and (17a) for the exchange hole:

(47) $\quad n^\mu(\mathbf{r})n^\nu(\mathbf{r}')n^\lambda(\mathbf{r}'') \cdot \underbrace{\tilde{g}_3(\mathbf{r},\mathbf{r}',\mathbf{r}'') \cdot n_\lambda(\mathbf{r}'')}_{\tilde{n}_{3,\lambda}(\mathbf{r},\mathbf{r}',\mathbf{r}'')} \quad = \quad \tilde{H}^{\mu\nu\lambda}(\mathbf{r},\mathbf{r}',\mathbf{r}'') \cdot n_\lambda(\mathbf{r}'')$

$\tilde{n}_{3,\lambda}$ is a four-vector with the dimension (volume)$^{-1}$. This allows a possible sum rule to be written in covariant form, similar to charge conservation for a conserved four-current:

$$\int \tilde{n}_{3,\lambda}(\mathbf{r},\mathbf{r}',\mathbf{r}'')\, d^3\mathbf{r}'' \;=\; \int \tilde{n}_{3,\lambda}(\mathbf{r},\mathbf{r}',\mathbf{r}'')\, d^4x''_\mu \;=\; 0 \quad \text{for all } \mathbf{r},\mathbf{r}'$$

The integral $d^4x''_\mu$ is taken over a time-like hypersurface in space-time, such as $d^3\mathbf{r}''$. One could also define a two-fermion density $\tilde{n}_3^{\mu\nu}(\mathbf{r},\mathbf{r}',\mathbf{r}'')$ for the exchange exciton by contracting (45) with $n_\nu(\mathbf{r}')n_\lambda(\mathbf{r}'')$. That produces a tensor with the dimension (volume)$^{-2}$:

(48) $\quad \tilde{n}_3^{\mu\nu}(\mathbf{r},\mathbf{r}',\mathbf{r}'') \;=\; \tilde{g}_3(\mathbf{r},\mathbf{r}',\mathbf{r}'') \cdot n^\nu(\mathbf{r}')n^\mu(\mathbf{r}'') \qquad \tilde{n}_{3,\nu\lambda}(\mathbf{r},\mathbf{r}',\mathbf{r}'') \;=\; \tilde{g}_3(\mathbf{r},\mathbf{r}',\mathbf{r}'') \cdot n_\nu(\mathbf{r}')n_\lambda(\mathbf{r}'')$

(49) $\quad n^\mu(\mathbf{r})n^\nu(\mathbf{r}')n^\lambda(\mathbf{r}'') \cdot \underbrace{\tilde{g}_3(\mathbf{r},\mathbf{r}',\mathbf{r}'') \cdot n_\nu(\mathbf{r}')n_\lambda(\mathbf{r}'')}_{\tilde{n}_{3,\nu\lambda}(\mathbf{r},\mathbf{r}',\mathbf{r}'')} \;=\; \tilde{H}^{\mu\nu\lambda}(\mathbf{r},\mathbf{r}',\mathbf{r}'') \cdot n_\nu(\mathbf{r}')n_\lambda(\mathbf{r}'')$

These definitions can be applied to the Fermi sea directly, as done in Section 4.2. For the Dirac sea one runs into the problem of defining the one-electron density $n^\mu(\mathbf{r})$, as discussed at the beginning of Section 2.3. Therefore we use the same cure as in (11a), where the total electron density is replaced by that of the reference electron: $n^\mu(\mathbf{r}) \to n_0^\mu(\mathbf{r}) = s_e \cdot \bar{\psi}_0(\mathbf{r})\gamma^\mu\psi_0(\mathbf{r})$. The reference electron used in (21a) produces non-vanishing components of $n^\mu, n^\nu, n^\lambda$ only for $\mu=\nu=\lambda=0$. Consequently the $4^3$ components of the 3$^{rd}$ rank tensors in (47),(49) collapse into one. That makes the implicit definition of the exchange exciton explicit.

Since the Dirac sea is homogeneous, the coordinates $\mathbf{r},\mathbf{r}',\mathbf{r}''$ appear in the form $(\mathbf{r}-\mathbf{r}'),(\mathbf{r}'-\mathbf{r}''),(\mathbf{r}''-\mathbf{r})$. One of them is redundant. For example, one can set the coordinate $\mathbf{r}$ of the reference electron to zero. Another option is the center-of-mass frame of the exchange exciton. This frame plays a special role in two-fermion Dirac equations [13] which would represent the next level of sophistication in describing the exchange exciton. Other possibilities would be the center-of-mass frame of the three fermions or one of the frames used to describe the negative ion of positronium [14].

The calculation of the three-fermion correlation amounts to the solution of a relativistic three-body problem – a difficult task. Two avenues for handling it are outlined in the following. One is based on a direct calculation of the expression (47) for the exchange exciton in Appendix C, the other uses the negative ion of positronium as model.

A closer look at the product of Slater determinants in (41),(44) reveals the following structure: Compared to the 4 terms from the pair wave function one has now 36 terms, 6 from $\psi_{ijk}$ multiplied by 6 from from $\bar\psi_{ijk}$. A single exchange $\psi_i \leftrightarrow \psi_j$, $\bar\psi_i \leftrightarrow \bar\psi_j$ generates two terms of the form $(ii \cdot jj) \to (i\widehat{j \cdot j}i)$ and $(ii \cdot jj) \to (\widehat{ji \cdot ij})$. They are complex conjugates. A double exchange of the form $(ii \cdot jj \cdot kk) \to (ij \cdot \widehat{ki \cdot j}k)$ involves all three wave functions. This leads to a classification of the terms by the number of exchanges:

A) $\quad +\tilde{\Sigma}_{i,j,k}[(ii)(jj)(kk)+(jj)(ii)(kk)]$ $\qquad$ No exchange (Coulomb) $\qquad$ (6 terms)

B) $\quad -\tilde{\Sigma}_{i,j,k}[(ij \cdot ji) + (ji \cdot ij)] \cdot (kk)$ $\qquad$ Single exchange $\qquad$ (18 terms)

C) $\quad +\tilde{\Sigma}_{i,j,k}[(ik \cdot ji \cdot kj)+(ki \cdot ij \cdot jk)]$ $\qquad$ Double exchange $\qquad$ (12 terms)

Each exchange produces a minus sign, such that the single exchange terms form the exchange hole while the double exchange terms produce the exchange electron. The terms specified above are to be combined with the 3 even permutations of i,j,k. Altogether one obtains 36 terms, as expected from the product of the two Slater determinants. If one of the indices i,j,k is assigned to the reference electron (labeled 0), the single exchange terms B) split into two categories:

B1)  $-\Sigma_{i,j}[(ij \cdot ji) + (ji \cdot ij)] \cdot (00)$    Single exchange · (00)    (6 terms)

B2)  $-\Sigma_{i,j}[(j0 \cdot 0j) + (0j \cdot j0)] \cdot (ii)$    Single exchange · (ii)    (6 terms)

   $-\Sigma_{i,j}[(i0 \cdot 0i) + (0i \cdot i0)] \cdot (jj)$    Single exchange · (jj)    (6 terms)

These terms are calculated and discussed in Appendix C after replacing i,j,k with **p**,**p**′,**p**″. The results still suffer from a divergence at small r during the evaluation of the sum rule for the exchange hole. To obtain finite results it would be desirable to replace the *ad-hoc* assignment of $n^\mu(\mathbf{r})$ to $n_0^\mu(\mathbf{r})$ by techniques from quantum electrodynamics, such as normal ordering of creation and annihilation operators and renormalization.

The second avenue for handling the three-fermion problem exploits analogies to the negative ion of positronium. It consists of two electrons and a positron in a triangular configuration, with equal bond lengths between the positron and the two electrons [14]. A similar arrangement can be inferred for the three-fermion system considered here by identifying the bond lengths with the distances obtained in (24),(27) from the three pair correlations between the reference electron $e_r$, the exchange hole $h_x$, and the exchange electron $e_x$:

(50)    $e_r - h_x \approx 1.3\,\lambda_C$    $e_r - e_x \approx 2.0\,\lambda_C$    $\theta(h_x) \approx 103.5°$    $e_r + h_x + e_x$

(51)    $e^- - e^+ \approx 5.5\,a_0$    $e^- - e^- \approx 9.0\,a_0$    $\theta(e^+) \approx 110°$    $e^- + e^+ + e^-$

The bond angle at the exchange hole is determined from the bond lengths via $\theta(h_x) = 2 \cdot \arcsin(\pi/4)$. The two equal bond lengths connected to the exchange hole ($e_r - h_x$, $e^- - h_x$) are 580 times shorter than the $e^- - e^+$ bond length. This demonstrates once more the strength of the exchange interaction in the Dirac sea compared to the Coulomb interaction in negative positronium. Closer scrutiny exposes several weak points in the analogy: 1) the exchange hole and exchange electron are many-body objects with a finite size, while electron and positron are point-like fundamental particles, 2) the reference and exchange electrons have parallel spins, while the two electrons in negative positronium have opposite spins, 3) the reference electron and the exchange electron are formed by negative energy states. Nevertheless, the same methods are applicable for handling these relativistic three-fermion problems.

### 4.2 Application to Solids

The concept of an exchange exciton could have applications beyond its original goal. For example, it might be useful in for characterizing exchange-correlation effects in semiconductors and insulators. Therefore we briefly consider the structure of the exchange exciton in the Fermi sea. Even though this is a metallic system, it can serve as a simple test case for exploring the overall structure of the three-fermion correlation. The pair correlation $g^F$ in the Fermi sea has already been discussed in Section 2.6. For the corresponding three-electron correlation $g_3^F$ one needs the three-electron densities $G^{000}$ from (44) and A),B),C), summed over all three momenta $\mathbf{p},\mathbf{p'},\mathbf{p''}$ up to $p_F$ (with $p_F \ll m_e$):

(52) $\quad A = +\tfrac{1}{6} \iiint d^3\mathbf{p}\, d^3\mathbf{p'}\, d^3\mathbf{p''}$ + permutations $\quad y = p_F \cdot |\mathbf{r}| \quad y' = p_F \cdot |\mathbf{r'}| \quad y'' = p_F \cdot |\mathbf{r''}|$

$\qquad \propto + n(y) \cdot n(y') \cdot n(y'') \qquad\qquad\qquad\qquad z = p_F \cdot |\mathbf{r'} - \mathbf{r''}| \quad z' = p_F \cdot |\mathbf{r''} - \mathbf{r}| \quad z'' = p_F \cdot |\mathbf{r} - \mathbf{r'}|$

(53) $\quad B = -\tfrac{1}{6} \iiint d^3\mathbf{p}\, d^3\mathbf{p'}\, d^3\mathbf{p''}\, \cos[\mathbf{p}(\mathbf{r'}-\mathbf{r})] \cdot \cos[\mathbf{p'}(\mathbf{r}-\mathbf{r'})] \cdot \cos[\mathbf{p''}(\mathbf{r''}-\mathbf{r''})]$ + permutations

$\qquad \propto - \{ |n_x^F(z'')|^2 \cdot n(y'') + |n_x^F(z)|^2 \cdot n(y) + |n_x^F(z')|^2 \cdot n(y') \}$

(54) $\quad C = +\tfrac{1}{6} \iiint d^3\mathbf{p}\, d^3\mathbf{p'}\, d^3\mathbf{p''}\, \cos[\mathbf{p}(\mathbf{r}-\mathbf{r'})] \cdot \cos[\mathbf{p'}(\mathbf{r'}-\mathbf{r''})] \cdot \cos[\mathbf{p''}(\mathbf{r''}-\mathbf{r})]$ + permutations

$\qquad \propto + |n_x^F(z'')| \cdot |n_x^F(z)| \cdot |n_x^F(z')|$

The Coulomb term A produces simply the product of three (constant) one-electron densities $n(\mathbf{r}),n(\mathbf{r'}),n(\mathbf{r''})$. The exchange hole B becomes equal to the standard result $g^F(r)$ in (35) after including the normalization factors for obtaining the dimension (volume)$^{-1}$ and satisfying the sum rule. The pairwise coordinate differences $(\mathbf{r}-\mathbf{r'}),(\mathbf{r'}-\mathbf{r''}),(\mathbf{r''}-\mathbf{r})$ each acquire their own exchange hole. The three exchange electrons are obtained from the double-exchange terms C. They have the form $|n_x^F|$ from (34) and appear as a product. It is gratifying to see that the exchange hole is now identical to its definition in solid state physics. The radii of the exchange hole and electron in (35) and (34) cannot be quantified by their expectation values (as in (24) and (27) for the Dirac sea), since the corresponding integrals diverge at large $r$. A possible measure is the half-height of the electron densities:

(55) $\quad g^F(r) = \tfrac{1}{2} \cdot g^F(0) \quad$ at $\quad r \approx 1.81/p_F \approx 0.29 \cdot \lambda_F$

(56) $\quad n_x^F(r) = \tfrac{1}{2} \cdot n_x^F(0) \quad$ at $\quad r \approx 2.50/p_F \approx 0.40 \cdot \lambda_F$

Furthermore, the undamped oscillations of the integral for the sum rule in (34) mask the true shape of the exchange electron. This problem extends even to small $r$, where the exchange electron dominates over the exchange hole. The exclusion principle requires the opposite, i.e., the vacuum electrons pushed away by the reference electron should be outside the exchange hole. These difficulties are due to long-range correlations in the Fermi sea induced by the sharp cutoff of the momentum integrals at $p_F$. They should not be a problem for semiconductors and insulators whose orbitals decay exponentially (compare the calculations of the exchange hole in [5]).

## 5. Summary and Outlook

This work stresses the exchange interaction of an electron with the Dirac sea as crucial ingredient for explaining its internal stability. The Coulomb explosion problem of classical electrodynamics is cured by the quantum-mechanical exchange interaction. It prevents direct self-interaction via the cancellation between the self-Coulomb and self-exchange terms. Instead, the exchange interaction of an electron with the Dirac sea provides a strong indirect self-interaction. That will make an important contribution to the force density balance which characterizes the stability of the electron. Guided by the analogy to the exchange hole in the Fermi sea and by Weisskopf's pioneering work with the Dirac sea, the exchange interaction of an electron with the Dirac sea is characterized.

The results can be summarized as follows:

1) An electron couples to the electrons and positrons in the vacuum of quantum electrodynamics predominantly via the exchange interaction.

2) The effect of the exchange interaction can be captured by generalizing the concept of an exchange hole from condensed matter physics to quantum electrodynamics. That leads to the addition of an exchange electron.

3) The exchange electron is necessary to maintain the charge neutrality of the vacuum. Exchange hole and electron form an exchange exciton in the singlet state. Thereby the angular momentum is conserved as well.

4) While a 3-fermion correlation is needed to generalize the definition of the exchange hole, an approximation by three pair correlations is more intuitive and easier to handle. Analytic results are obtained for the charge densities and the average distances between the exchange hole, the exchange electron, and the reference electron. The bond lengths suggest a triangular arrangement similar to that in the negative ion of positronium.

5) A calculation of the electrostatic force densities generated by the exchange interaction makes a first step toward establishing a force density balance for the electron. Judging from a previous study of force densities in the hydrogen atom, a complete force balance will require a calculation of the confinement force density.

6) It would be desirable to develop the heuristic approach adopted here more rigorously by borrowing techniques from quantum electrodynamics. In particular, Wick's theorem could be used to decompose the 3-fermion correlation into 2-fermion correlations [16].

7) The concept of an exchange exciton might have applications in solid state physics for characterizing the exchange interaction in insulators and semiconductors. The neutral exchange exciton is more realistic than the charged exchange hole in that case, because the electron displaced by the exchange hole cannot delocalize.

**Appendix A: Units, Notation, Fourier Transforms**

The units are $\hbar, c, m_e$, with the length unit $\lambdabar_C = \hbar/m_e c$ (reduced Compton wavelength). The metric is $g^{\mu\nu} = (+ - - -) \cdot \delta^{\mu\nu}$, with summation over pairs of Greek indices and three-vectors in bold. The Gaussian system is used for electromagnetism ($e > 0$). The Fourier transform $\tilde{A}(\mathbf{p})$ of $A(\mathbf{r})$ and its inverse are defined as follows (see also Appendix D):

(A1) $\quad \tilde{A}(p^0) = \int A(t) \exp(+i p^0 t)\, dt \qquad\qquad A(t) = (2\pi)^{-1} \cdot \int \tilde{A}(p^0) \exp(-i p^0 t)\, dp^0$

(A2) $\quad \tilde{A}(\mathbf{p}) = \int A(\mathbf{r}) \exp(-i\mathbf{p}\mathbf{r})\, d^3 r \qquad A(\mathbf{r}) = (2\pi)^{-3} \cdot \int \tilde{A}(\mathbf{p}) \exp(+i\mathbf{p}\mathbf{r})\, d^3 p$

(A3) $\quad \tilde{A}(\mathbf{p}) = 1 \qquad\qquad\qquad\qquad\qquad A(\mathbf{r}) = \delta^3(\mathbf{r}) = (2\pi)^{-3} \cdot \int \exp(i\mathbf{p}\mathbf{r})\, d^3 p$

(A4) $\quad \tilde{A}(\mathbf{p}) = \delta^3(\mathbf{p}) \qquad\qquad\qquad\quad\ A(\mathbf{r}) = (2\pi)^{-3}$

**Appendix B: Calculations Involving Dirac Spinors**

The wave functions $\psi(\mathbf{r},t)$ defined in (20a) contain the Dirac spinors $u(\mathbf{p},s), \hat{u}(\mathbf{p},s)$ for electrons and $v(\mathbf{p},s), \hat{v}(\mathbf{p},s)$ for positrons [12] together with the plane waves given in (B3). Some of their properties are listed in the following (see also (20b,c)).

(B1) $\quad u(\mathbf{p},\uparrow) = [(E+m_e)/2E]^{1/2} \cdot \left\{ 1, 0, \dfrac{p_z}{E+m_e}, \dfrac{p_x + i p_y}{E+m_e} \right\} \qquad E = |p^0| = (m_e^2 + \mathbf{p}^2)^{1/2}$

$\quad\ \ u(\mathbf{p},\downarrow) = [(E+m_e)/2E]^{1/2} \cdot \left\{ 0, 1, \dfrac{p_x - i p_y}{E+m_e}, \dfrac{-p_z}{E+m_e} \right\}$

$p^0 < 0 \begin{cases} \hat{u}(\mathbf{p},\uparrow) = [(E+m_e)/2E]^{1/2} \cdot \left\{ \dfrac{-p_z}{E+m_e}, -\dfrac{p_x + i p_y}{E+m_e}, 1, 0 \right\} \\[1em] \hat{u}(\mathbf{p},\downarrow) = [(E+m_e)/2E]^{1/2} \cdot \left\{ -\dfrac{p_x - i p_y}{E+m_e}, \dfrac{p_z}{E+m_e}, 0, 1 \right\} \end{cases}$

$\qquad\qquad\qquad\qquad\qquad\qquad\qquad\qquad\qquad\qquad\qquad\qquad\qquad p^\mu \leftrightarrow -p^\mu,\ \uparrow \leftrightarrow \downarrow,\ e^- \leftrightarrow e^+$

(B2) $\quad v(\mathbf{p},\uparrow) = [(E+m_e)/2E]^{1/2} \cdot \left\{ \dfrac{p_x - i p_y}{E+m_e}, \dfrac{-p_z}{E+m_e}, 0, 1 \right\}$

$\quad\ \ v(\mathbf{p},\downarrow) = [(E+m_e)/2E]^{1/2} \cdot \left\{ \dfrac{p_z}{E+m_e}, \dfrac{p_x + i p_y}{E+m_e}, 1, 0 \right\}$

$p^0 < 0 \begin{cases} \hat{v}(\mathbf{p},\uparrow) = [(E+m_e)/2E]^{1/2} \cdot \left\{ 0, 1, -\dfrac{p_x - i p_y}{E+m_e}, \dfrac{p_z}{E+m_e} \right\} \\[1em] \hat{v}(\mathbf{p},\downarrow) = [(E+m_e)/2E]^{1/2} \cdot \left\{ 1, 0, \dfrac{-p_z}{E+m_e}, -\dfrac{p_x + i p_y}{E+m_e} \right\} \end{cases}$

Dirac equation for electrons and positrons:

(B3) $\quad (+\gamma^\mu p_\mu - m_e) \cdot u(\mathbf{p},s) = 0 \qquad (i\gamma^\mu \partial_\mu - m_e)[u(\mathbf{p},s) \exp(-i p_\mu x^\mu)] = 0 \qquad p^0 > 0$

$\qquad\ (-\gamma^\mu p_\mu - m_e) \cdot \hat{u}(\mathbf{p},s) = 0 \qquad (i\gamma^\mu \partial_\mu - m_e)[\hat{u}(\mathbf{p},s) \exp(+i p_\mu x^\mu)] = 0 \qquad p^0 < 0$

$\qquad\ (-\gamma^\mu p_\mu - m_e) \cdot v(\mathbf{p},s) = 0 \qquad (i\gamma^\mu \partial_\mu - m_e)[v(\mathbf{p},s) \exp(+i p_\mu x^\mu)] = 0 \qquad p^0 > 0$

$\qquad\ (+\gamma^\mu p_\mu - m_e) \cdot \hat{v}(\mathbf{p},s) = 0 \qquad (i\gamma^\mu \partial_\mu - m_e)[\hat{v}(\mathbf{p},s) \exp(-i p_\mu x^\mu)] = 0 \qquad p^0 < 0$

Charge conjugation:

(B4)    $i\gamma^2 u(\mathbf{p},\uparrow)^* = +\hat{u}(-\mathbf{p},\downarrow) = +v(\mathbf{p},\uparrow)$      $i\gamma^2 \hat{u}(\mathbf{p},\uparrow)^* = -u(-\mathbf{p},\downarrow) = -\hat{v}(\mathbf{p},\uparrow)$

       $i\gamma^2 u(\mathbf{p},\downarrow)^* = -\hat{u}(-\mathbf{p},\uparrow) = -v(\mathbf{p},\downarrow)$      $i\gamma^2 \hat{u}(\mathbf{p},\downarrow)^* = +u(-\mathbf{p},\uparrow) = +\hat{v}(\mathbf{p},\downarrow)$

Scalar density and completeness:

(B5a)    $\bar{u}(\mathbf{p},s)\cdot u(\mathbf{p},s') = +m_e/E \cdot \delta_{s,s'}$      $\bar{v}(\mathbf{p},s)\cdot v(\mathbf{p},s') = -m_e/E \cdot \delta_{s,s'}$

       $\bar{\hat{u}}(\mathbf{p},s)\cdot \hat{u}(\mathbf{p},s') = -m_e/E \cdot \delta_{s,s'}$      $\bar{\hat{v}}(\mathbf{p},s)\cdot \hat{v}(\mathbf{p},s') = +m_e/E \cdot \delta_{s,s'}$

       $\Sigma_s u(\mathbf{p},s)_\alpha \bar{u}(\mathbf{p},s)_\beta = +(\gamma^\mu_{\alpha\beta}p_\mu + m_e\delta_{\alpha\beta})/2E$      $\Sigma_s v(\mathbf{p},s)_\alpha \bar{v}(\mathbf{p},s)_\beta = +(\gamma^\mu_{\alpha\beta}p_\mu - m_e\delta_{\alpha\beta})/2E$

       $\Sigma_s \hat{u}(\mathbf{p},s)_\alpha \bar{\hat{u}}(\mathbf{p},s)_\beta = -(\gamma^\mu_{\alpha\beta}p_\mu + m_e\delta_{\alpha\beta})/2E$      $\Sigma_s \hat{v}(\mathbf{p},s)_\alpha \bar{\hat{v}}(\mathbf{p},s)_\beta = -(\gamma^\mu_{\alpha\beta}p_\mu - m_e\delta_{\alpha\beta})/2E$

Particle density and orthonormality:

(B5b)    $\bar{u}(\mathbf{p},s)\gamma^0 u(\mathbf{p},s') = +\delta_{s,s'}$      $\bar{v}(\mathbf{p},s)\gamma^0 v(\mathbf{p},s') = +\delta_{s,s'}$

       $\bar{\hat{u}}(\mathbf{p},s)\gamma^0 \hat{u}(\mathbf{p},s') = +\delta_{s,s'}$      $\bar{\hat{v}}(\mathbf{p},s)\gamma^0 \hat{v}(\mathbf{p},s') = +\delta_{s,s'}$

       $\bar{u}(\mathbf{p},s)\gamma^0 \hat{u}(\mathbf{p},s') = 0$      $\bar{v}(\mathbf{p},s)\gamma^0 \hat{v}(\mathbf{p},s') = 0$

Current density:

(B6)    $\bar{u}(\mathbf{p},s)\,\gamma\, u(\mathbf{p},s') = +\mathbf{p}/E \cdot \delta_{s,s'}$      $\bar{v}(\mathbf{p},s)\,\gamma\, v(\mathbf{p},s') = +\mathbf{p}/E \cdot \delta_{s,s'}$

       $\bar{\hat{u}}(\mathbf{p},s)\,\gamma\,\hat{u}(\mathbf{p},s') = -\mathbf{p}/E \cdot \delta_{s,s'}$      $\bar{\hat{v}}(\mathbf{p},s)\,\gamma\,\hat{v}(\mathbf{p},s') = -\mathbf{p}/E \cdot \delta_{s,s'}$

**Biquadratic exchange terms with one p-integration**

Vector ⊗ vector (charge/current density):

(B7a)    $-\tfrac{1}{2}[+\int d^3(\mathbf{p}/m_e)\Sigma_s \bar{u}(\mathbf{p},s)\gamma^\mu \hat{u}(\mathbf{0},\uparrow)\cdot\bar{\hat{u}}(\mathbf{0},\uparrow)\gamma^\nu u(\mathbf{p},s)\cdot e^{-i(\mathbf{pr}-\mathbf{pr'})} + cc$

       $-\int d^3(\mathbf{p}/m_e)\Sigma_s \bar{v}(\mathbf{p},s)\gamma^\mu \hat{u}(\mathbf{0},\uparrow)\cdot\bar{\hat{u}}(\mathbf{0},\uparrow)\gamma^\nu \hat{v}(\mathbf{p},s)\cdot e^{+i(\mathbf{pr}-\mathbf{pr'})} + cc\,]$

   $= -g^{\mu\nu}\cdot\int \tfrac{1}{2}\{[1+\tfrac{m_e}{E(\mathbf{p})}]-[1-\tfrac{m_e}{E(\mathbf{p})}]\}\cdot\cos[\mathbf{p}(\mathbf{r}-\mathbf{r'})]\,d^3(\mathbf{p}/m_e)$

   $= -g^{\mu\nu}\cdot\int \tfrac{m_e}{E(\mathbf{p})}\cos[\mathbf{p}(\mathbf{r}-\mathbf{r'})]\,d^3(\mathbf{p}/m_e) = \tilde{H}^{\mu\nu}(\mathbf{r},\mathbf{r'})$      (after averaging over $^{\mu\nu}$ and $^{\nu\mu}$)

Scalar ⊗ scalar (mass density):

(B7b)    $-\tfrac{1}{2}[+\int d^3(\mathbf{p}/m_e)\Sigma_s \bar{u}(\mathbf{p},s)\hat{u}(\mathbf{0},\uparrow)\cdot\bar{\hat{u}}(\mathbf{0},\uparrow)u(\mathbf{p},s)\cdot e^{-i(\mathbf{pr}-\mathbf{pr'})} + cc$

       $-\int d^3(\mathbf{p}/m_e)\Sigma_s \bar{v}(\mathbf{p},s)\hat{u}(\mathbf{0},\uparrow)\cdot\bar{\hat{u}}(\mathbf{0},\uparrow)\hat{v}(\mathbf{p},s)\cdot e^{+i(\mathbf{pr}-\mathbf{pr'})} + cc\,]$

   $= -1\cdot\int \tfrac{m_e}{E(\mathbf{p})}\cos[\mathbf{p}(\mathbf{r}-\mathbf{r'})]\,d^3(\mathbf{p}/m_e)$

Tensor ⊗ tensor (dipole density):

(B7c)    $-\tfrac{1}{2}[+\int d^3(\mathbf{p}/m_e)\Sigma_s \bar{u}(\mathbf{p},s)\sigma^{\mu\nu}\hat{u}(\mathbf{0},\uparrow)\cdot\bar{\hat{u}}(\mathbf{0},\uparrow)\sigma^{\rho\sigma}u(\mathbf{p},s)\cdot e^{-i(\mathbf{pr}-\mathbf{pr'})} + cc$

       $-\int d^3(\mathbf{p}/m_e)\Sigma_s \bar{v}(\mathbf{p},s)\sigma^{\mu\nu}\hat{u}(\mathbf{0},\uparrow)\cdot\bar{\hat{u}}(\mathbf{0},\uparrow)\sigma^{\rho\sigma}\hat{v}(\mathbf{p},s)\cdot e^{+i(\mathbf{pr}-\mathbf{pr'})} + cc\,]$

   $= -g^{\mu\rho}g^{\nu\sigma}\cdot\int \tfrac{m_e}{E(\mathbf{p})}\cos[\mathbf{p}(\mathbf{r}-\mathbf{r'})]\,d^3(\mathbf{p}/m_e)$      (after averaging over $^{\mu\nu\rho\sigma}$ and $^{\rho\sigma\mu\nu}$)

The contributions from the electron and positron Dirac seas are given for the vector exchange term in (B7a). Both contribute equally to the exchange hole, but they also generate δ-functions which compensate each other.

**Biquadratic exchange terms with two p-integrations**

$\hat{u}(\mathbf{0},\uparrow)$ in (B7a) is generalized to $\hat{u}(\mathbf{p}',s'), \hat{v}(\mathbf{p}',s')$ with an extra summation over $\hat{u},\hat{v}$ and $\mathbf{p}',s'$:

(B8) $\quad -\tfrac{1}{2}[+\iint d^3\mathbf{p}\, d^3\mathbf{p}'\, \Sigma_s \Sigma_{s'}\, \bar{\hat{u}}(\mathbf{p},s)\gamma^\mu \hat{u}(\mathbf{p}',s') \cdot \bar{\hat{u}}(\mathbf{p}',s')\gamma^\nu \hat{u}(\mathbf{p},s) \cdot e^{j(-\mathbf{p}\,\mathbf{r}+\mathbf{p}'\mathbf{r}-\mathbf{p}'\mathbf{r}'+\mathbf{p}\,\mathbf{r}')} + \text{cc}$

$\quad\quad +\iint d^3\mathbf{p}\, d^3\mathbf{p}'\, \Sigma_s \Sigma_{s'}\, \bar{\hat{v}}(\mathbf{p},s)\gamma^\mu \hat{v}(\mathbf{p}',s') \cdot \bar{\hat{v}}(\mathbf{p}',s')\gamma^\nu \hat{v}(\mathbf{p},s) \cdot e^{j(+\mathbf{p}\,\mathbf{r}-\mathbf{p}'\mathbf{r}+\mathbf{p}'\mathbf{r}'-\mathbf{p}\,\mathbf{r}')} + \text{cc}$

$\quad\quad -\iint d^3\mathbf{p}\, d^3\mathbf{p}'\, \Sigma_s \Sigma_{s'}\, \bar{\hat{u}}(\mathbf{p},s)\gamma^\mu \hat{v}(\mathbf{p}',s') \cdot \bar{\hat{v}}(\mathbf{p}',s')\gamma^\nu \hat{u}(\mathbf{p},s) \cdot e^{j(-\mathbf{p}\,\mathbf{r}-\mathbf{p}'\mathbf{r}+\mathbf{p}'\mathbf{r}'+\mathbf{p}\,\mathbf{r}')} + \text{cc}$

$\quad\quad -\iint d^3\mathbf{p}\, d^3\mathbf{p}'\, \Sigma_s \Sigma_{s'}\, \bar{\hat{v}}(\mathbf{p},s)\gamma^\mu \hat{u}(\mathbf{p}',s') \cdot \bar{\hat{u}}(\mathbf{p}',s')\gamma^\nu \hat{v}(\mathbf{p},s) \cdot e^{j(+\mathbf{p}\,\mathbf{r}+\mathbf{p}'\mathbf{r}-\mathbf{p}'\mathbf{r}'-\mathbf{p}\,\mathbf{r}')} + \text{cc}\,]$

The exchange hole is determined by the component $\mu=0, \nu=0$, which takes the form:

(B9) $\quad -2\iint d^3\mathbf{p}\, d^3\mathbf{p}'\, [\,1/(E\cdot E') + 1 + \mathbf{p}\cdot\mathbf{p}'/(E\cdot E')\,]\cdot\cos[(\mathbf{p}-\mathbf{p}')(\mathbf{r}-\mathbf{r}')]\quad\quad m_e=1\quad E=(1+\mathbf{p}^2)^{1/2}$

$\quad\quad -2\iint d^3\mathbf{p}\, d^3\mathbf{p}'\, [\,1/(E\cdot E') - 1 - \mathbf{p}\cdot\mathbf{p}'/(E\cdot E')\,]\cdot\cos[(\mathbf{p}+\mathbf{p}')(\mathbf{r}-\mathbf{r}')]\quad\quad\quad\quad\quad\quad E'=(1+\mathbf{p}'^2)^{1/2}$

The first line originates from the terms $+(e^-e^- + e^+e^+)$, the second from $-(e^-e^+ + e^+e^-)$. To separate the integrals over $\mathbf{p},\mathbf{p}'$ the $\cos[(\mathbf{p}\pm\mathbf{p}')(\mathbf{r}-\mathbf{r}')]$ factors are expanded into products of sin and cos of $\mathbf{p},\mathbf{p}'$. A rearrangement of the terms in (B9) then yields (with $r'=|\mathbf{r}-\mathbf{r}'|$):

(B10) $\quad -4\iint d^3\mathbf{p}\, d^3\mathbf{p}'\, \cos[\mathbf{p}(\mathbf{r}-\mathbf{r}')]\cdot\cos[\mathbf{p}'(\mathbf{r}-\mathbf{r}')]/(E\cdot E')\quad\quad = -4[1/2\pi^2 \cdot r'^{-1} K_1(r')]^2$

$\quad\quad -4\iint d^3\mathbf{p}\, d^3\mathbf{p}'\, \sin[\mathbf{p}(\mathbf{r}-\mathbf{r}')]\cdot\sin[\mathbf{p}'(\mathbf{r}-\mathbf{r}')] = 0$

$\quad\quad -4\iint d^3\mathbf{p}\, d^3\mathbf{p}'\, \sin[\mathbf{p}(\mathbf{r}-\mathbf{r}')]\cdot\sin[\mathbf{p}'(\mathbf{r}-\mathbf{r}')]\cdot\mathbf{p}\cdot\mathbf{p}'/(E\cdot E') = -4[1/2\pi^2 \cdot r'^{-1} K_2(r')]^2$

The second integral vanishes, because the integrand has odd inversion symmetry. The first integral produces the square of the exchange hole according to (B7a),(D5). The third integral is the squared gradient of the exchange hole according to (D8). Analogous results are obtained for $r''=|\mathbf{r}-\mathbf{r}''|$.

**Appendix C: Three-Fermion Correlations**

**A) Coulomb terms**

The Coulomb terms are triple products of the one-electron densities $(ii)=n^\mu$ in (11a),(21a) which are assigned by definition to $\delta^{\mu 0}\cdot(2\pi\lambda_C)^{-3}$. The 6 permutations of the three indices i,j,k produce a factor 6 on the left side of (47). After moving it to the right side one obtains the normalization factor $\tfrac{1}{6}$ in the definition of the exchange terms, such as (B12).

**B1) Biquadratic exchange terms with the factor (00)**

These biquadratic products of the form (ij·ji) with two **p**-integrations are calculated in (B8)-(B10). The additional factor $(00)=\delta^{\mu 0}\cdot(2\pi\lambda_C)^{-3}$ generates the prefactor $(2\pi)^{-3}$ for the second inverse Fourier transform. The evaluation in (B10) yields the square $(\tilde{n}_x^h)^2$ of exchange hole obtained from the pair density in (23), plus an additional term from the square of its gradient $(\nabla\tilde{n}_x^h)\cdot(\nabla\tilde{n}_x^h)$. Both diverge at $\mathbf{r}=\mathbf{0}$ such that the charge of the exchange hole diverges. This ultraviolet divergence requires either a different definition of the exchange hole or appropriate renormalization before meaningful results can be obtained.

**B2) Biquadratic exchange terms with the factor (ii) or (jj)**

The biquadratic products of the form (j0·0j) with one **p**-integration are calculated in (B7a). But the factor (ii) vanishes, since electrons and positrons cancel each other. That leaves the 6 exchange terms from B1) to match the 6 Coulomb terms from A).

**C) Triquadratic double exchange terms**

The double exchange terms of the form (i0·ji·0j) involve three coordinates explicitly. The corresponding spinor products are:

(B12)  $\frac{1}{6}[+\iint d^3\mathbf{p}\,d^3\mathbf{p}'\,\Sigma_s\Sigma_{s'}\,\bar{u}(\mathbf{p},s)\gamma^\mu \hat{u}(\mathbf{0},\uparrow) \cdot \bar{u}(\mathbf{p}',s')\gamma^\nu \hat{u}(\mathbf{p},s) \cdot \bar{u}(\mathbf{0},\uparrow)\gamma^\lambda \hat{u}(\mathbf{p},s) \ + \ cc$

$+ \iint d^3\mathbf{p}\,d^3\mathbf{p}'\,\Sigma_s\Sigma_{s'}\,\bar{v}(\mathbf{p},s)\gamma^\mu \hat{u}(\mathbf{0},\uparrow) \cdot \bar{v}(\mathbf{p}',s')\gamma^\nu \hat{v}(\mathbf{p},s) \cdot \bar{u}(\mathbf{0},\uparrow)\gamma^\lambda \hat{v}(\mathbf{p},s) \ + \ cc$

$- \iint d^3\mathbf{p}\,d^3\mathbf{p}'\,\Sigma_s\Sigma_{s'}\,\bar{u}(\mathbf{p},s)\gamma^\mu \hat{u}(\mathbf{0},\uparrow) \cdot \bar{v}(\mathbf{p}',s')\gamma^\nu \hat{u}(\mathbf{p},s) \cdot \bar{u}(\mathbf{0},\uparrow)\gamma^\lambda \hat{v}(\mathbf{p},s) \ + \ cc$

$- \iint d^3\mathbf{p}\,d^3\mathbf{p}'\,\Sigma_s\Sigma_{s'}\,\bar{v}(\mathbf{p},s)\gamma^\mu \hat{u}(\mathbf{0},\uparrow) \cdot \bar{u}(\mathbf{p}',s')\gamma^\nu \hat{v}(\mathbf{p},s) \cdot \bar{u}(\mathbf{0},\uparrow)\gamma^\lambda \hat{u}(\mathbf{p},s) \ + \ cc \ ]$

The non-vanishing components $\mu=\nu=\lambda=0$ become (with $r'=|\mathbf{r}-\mathbf{r}'|$, $r''=|\mathbf{r}-\mathbf{r}''|$):

(B13)  $+\frac{1}{3}\iint d^3\mathbf{p}\,d^3\mathbf{p}'\,\cos[\mathbf{p}(\mathbf{r}-\mathbf{r}')]\cdot\cos[\mathbf{p}'(\mathbf{r}-\mathbf{r}'')]/(E\cdot E')$

$+\frac{1}{3}\iint d^3\mathbf{p}\,d^3\mathbf{p}'\,\sin[\mathbf{p}(\mathbf{r}-\mathbf{r}')]\cdot\sin[\mathbf{p}'(\mathbf{r}-\mathbf{r}'')] = 0$

$+\frac{1}{3}\iint d^3\mathbf{p}\,d^3\mathbf{p}'\,\sin[\mathbf{p}(\mathbf{r}-\mathbf{r}')]\cdot\sin[\mathbf{p}'(\mathbf{r}-\mathbf{r}'')]\cdot \mathbf{p}\cdot\mathbf{p}'/(E\cdot E')$

$= +\frac{1}{3}[1/2\pi^2 \cdot r'^{-1} K_1(r')]\cdot[1/2\pi^2 \cdot r''^{-1} K_1(r'')] + \frac{1}{3}[1/2\pi^2 \cdot r'^{-1} K_2(r')]\cdot[1/2\pi^2 \cdot r''^{-1} K_2(r'')]$

The other two even permutations of $\mathbf{r},\mathbf{r}',\mathbf{r}''$ have been omitted. The integrals are similar to those in (B10), except that there are two distances involved here. The plus sign of these double-exchange terms makes them candidates for the exchange electron.

**Appendix D:  Special Functions and Inverse Fourier Transforms**

Recursion relations for modified Bessel functions $K_n$ and Struve functions $\mathbf{L}_m$ [15]:

(D1)  $K_{n+1}(z) = 2n/z \cdot K_n(z) + K_{n-1}(z)$ $\qquad K'_n(z) = n/z \cdot K_n(z) - K_{n+1}(z)$

(D2)  $\mathbf{L}_{-3}(z) = -4/z \cdot \mathbf{L}_{-2}(z) + \mathbf{L}_{-1}(z) - 2/\pi \cdot z^{-2}$

$\mathbf{L}_{-2}(z) = -2/z \cdot \mathbf{L}_{-1}(z) + \mathbf{L}_0(z) + 2/\pi \cdot z^{-1}$

$\mathbf{L}_{-1}(z) = \mathbf{L}_1(z) + 2/\pi$ $\qquad \mathbf{L}'_{-n}(z) = n/z \cdot \mathbf{L}_{-n}(z) + \mathbf{L}_{-(n+1)}(z)$

(D3)  $Ki_1(z) = \int_z^\infty K_0(y)\,dy = \pi/2\,\{1 - z\cdot[K_0(z)\cdot\mathbf{L}_{-1}(z) + K_1(z)\cdot\mathbf{L}_0(z)]\}$

Weisskopf's pair correlation $\tilde{G}$ is related to the exchange hole density $|\tilde{n}_x^h|$ via:

(D4)  $i H_0^{(1)}(iz) = 2/\pi \cdot K_0(z)$ $\qquad K'_0(z) = -K_1(z)$

The inverse Fourier transform for obtaining the electron density in real space is defined in Appendix A. It is illustrated below for the absolute value of the exchange hole $|\tilde{n}_x^h(r)| = -\tilde{n}_x^h(r)$ in (22),(23) and the exchange electron $\tilde{n}_x^e(r)$ in (25),(26). First, the angular integrations are carried out using the variable $z=-\cos(\theta_{pr})$. Then radial p-integrand is integrated with respect to the parameter r. That converts $\sin(pr)$ into $-\partial/\partial r[p^{-1}\cos(pr)]$ and thereby improves the convergence of the p-integration for $p\to\infty$. The integral then can be carried out analytically. If the integrand is even in **p**, the integral over the imaginary part of $\exp(i\mathbf{pr})$ vanishes and leaves $\cos(\mathbf{pr})$. For example, the Fourier integral for the exchange hole has the form:

(D5) $\quad +(2\pi)^{-3}\cdot\int E(p)^{-1}\cdot\cos(\mathbf{pr})\,d^3p \qquad\qquad p=|\mathbf{p}|\quad r=|\mathbf{r}|\quad E(p)=(1+p^2)^{1/2}\quad m_e=1$

$\quad = +(2\pi)^{-2}\cdot\int_0^\infty E(p)^{-1} p^2\cdot\int_{-1}^{+1}\cos(pr\cdot z)\,dz\,dp \qquad \mathbf{pr}=pr\cdot\cos(\theta_{pr})=-pr\cdot z$

$\quad = +2(2\pi)^{-2}\cdot\int_0^\infty E(p)^{-1} p^2\cdot\sin(pr)/pr\,dp$

$\quad = -2(2\pi)^{-2}\cdot r^{-1}\partial/\partial r\int_0^\infty E(p)^{-1}\cdot\cos(pr)\,dp$

$\quad = -1/2\pi^2\cdot r^{-1}\partial/\partial r\,K_0(r)\ =\ +1/2\pi^2\cdot r^{-1}K_1(r)$

In practice one can use the following shortcut, as demonstrated for the exchange electron:

(D6) $\quad +(2\pi)^{-3}\cdot\int E(p)^{-2}\cdot\cos(\mathbf{pr})\,d^3p$

$\quad = -2(2\pi)^{-2}\cdot r^{-1}\partial/\partial r\int_0^\infty E(p)^{-2}\cdot\cos(pr)\,dp$

$\quad = -1/4\pi\cdot r^{-1}\partial/\partial r\,e^{-r}\ =\ +1/4\pi\cdot r^{-1}e^{-r}$

If the integrand is odd in **p**, the inverse Fourier transform involves the imaginary part $i\sin(\mathbf{pr})$ of $\exp(i\mathbf{pr})$. It has the general form $\tilde{\mathbf{A}}(\mathbf{p})=\mathbf{p}\cdot\tilde{A}(p)$, where $\tilde{A}(p)$ is a scalar and $\tilde{\mathbf{A}}(\mathbf{p})$ a vector. If $\tilde{\mathbf{A}}(\mathbf{p}),\tilde{A}(p)$ are real, their inverse transforms $\mathbf{A}(\mathbf{r}),A(r)$ become imaginary (and vice versa). The transform of the vector $\tilde{\mathbf{A}}(\mathbf{p})$ can be reduced to that of the scalar $\tilde{A}(p)$ by generating the factor **p** via the gradient operator $-i\partial/\partial\mathbf{r}$, applied to the scalar integrand (analogous to the definition of the momentum operator $\mathbf{P}=-i\partial/\partial\mathbf{r}$):

(D7) $\quad \mathbf{A}(\mathbf{r}) = (2\pi)^{-3}\cdot\int\tilde{\mathbf{A}}(\mathbf{p})\cdot\exp(i\mathbf{pr})\,d^3p$

$\quad\quad = (2\pi)^{-3}\cdot\int\mathbf{p}\cdot\tilde{A}(p)\cdot\exp(i\mathbf{pr})\,d^3p$

$\quad\quad = -i\partial/\partial\mathbf{r}[(2\pi)^{-3}\cdot\int\tilde{A}(p)\cdot\exp(i\mathbf{pr})\,d^3p]\ =\ -i\partial/\partial\mathbf{r}\,A(r)$

The calculations in Appendix B use the imaginary part of this equation to determine integrals of the type $\int\mathbf{p}\cdot\tilde{A}(p)\cdot\sin(\mathbf{pr})\,d^3p$. Such integrals appear in the calculation of the exchange hole from the three-fermion correlation:

(D8) $\quad +(2\pi)^{-3}\cdot\int\mathbf{p}\cdot E(p)^{-1}\cdot\sin(\mathbf{pr})\,d^3p$

$\quad = +(2\pi)^{-3}\cdot(-\partial/\partial\mathbf{r})\int E(p)^{-1}\cdot\cos(\mathbf{pr})\,d^3p \qquad$ using (D5)

$\quad = -\partial/\partial\mathbf{r}[1/2\pi^2\cdot r^{-1}K_1(r)]\ =\ +1/2\pi^2\cdot r^{-2}K_2(r)\cdot\mathbf{r}$

**References**

1. Electron self-interaction, early work: M. Abraham, *Classical Theory of Radiating Electrons*, Annalen der Physik **10**, 105 (1903); H. Poincaré, Comptes Rendus **140**, 1504 (1905); H. A. Lorentz, *The Theory of Electrons*, Columbia University Press, New York (1909), 2$^{nd}$ Edition, Dover, New York (1952); E. Fermi, *Über einen Widerspruch zwischen der Elektrodynamischen und der Relativistischen Theorie der Elektro-magnetischen Masse*, Physikalische Zeitschrift **23**, 340 (1922).
   Overviews: R. P. Feynman, *The Feynman Lectures on Physics*, Vol. II, Ch. 28 and his Nobel Lecture, *Nobel Prize in Physics 1963-1970*, Elsevier, Amsterdam (1972); P. Pearle, *Classical Electron Models*, Ch. 7 in: *Electromagnetism, Paths to Research*, ed. by D. Teplitz, Plenum Press, New York (1982); C. A. Brau, *Modern Problems in Classical Electrodynamics*, Oxford University Press, New York (2004).
2. V. F. Weisskopf, *On the Self-Energy and the Electromagnetic Field of the Electron*, Phys. Rev. **56**, 72 (1939); P. W. Milonni, *The Quantum Vacuum, an Introduction to Quantum Electrodynamics*, Academic Press, San Diego (1994), Ch. 11.4, 11.5. Notice that Weisskopf subtracted the electron density of the vacuum state from a state with an added reference electron. This method is not applicable here, since the reference electron is part of the Dirac sea (by analogy to the Fermi sea).
3. Early work on the exchange hole: J. C. Slater, *A Simplification of the Hartree-Fock Method*, Phys. Rev. **81**, 385 (1951); V. W. Maslen, *The Fermi, or Exchange, Hole in Atoms*, Proceedings of the Physical Society A **69**, 734 (1956); O. Gunnarson and B. I. Lundqvist, *Exchange and Correlation in Atoms, Molecules, and Solids by the Spin-Density-Functional Formalism*, Phys. Rev. B **13**, 4247 (1976).
4. Overview of the exchange hole: A. R. Williams and U. von Barth, *Applications of Density Functional Theory to Atoms, Molecules, and Solids*, Ch. 4 in: *Theory of the Inhomogeneous Electron Gas*, ed. by S. Lundqvist and N. H. March, Plenum Press, New York (1983); W. Kohn, *Nobel Lectures, Chemistry 1996-2000*, Editor I. Grenthe, World Scientific Publishing, Singapore (2003), p. 213; D. Pines and P. Nozières, *The Theory of Quantum Liquids*, Vol. I: *Normal Fermi Liquids*, Addison-Wesley (1989), Ch. 5. The symbols n, g, (g–1), G and the meaning of pair density, pair distribution, pair correlation, and exchange hole vary in the literature. A quick assessment of their usage can be obtained via the dimensionality, which ranges from (volume)$^0$ to (volume)$^{-3}$. An additional criterion is the number of momentum integrations. Also notice that a spin-polarized electron gas would require a non-trivial 2×2 matrix for the spin combinations of an electron pair. That is not necessary here, since the Dirac sea is unpolarized.
5. Exchange hole in insulators: S. Horsch, P. Horsch, P. Fulde, *Electronic Excitations in Semiconductors II Application of the Theory to Diamond*, Phys. Rev. B **29**, 1870 (1984); S. Fahy, X. Y. Wang, S. G. Louie, *Pair-Correlation Function and Single-Particle Occupation Numbers in Diamond and Silicon*, Phys. Rev. Lett. **65**, 1478 (1990).
6. Relativistic exchange hole: D. E. Ellis, *A Relativistic Exchange Potential*, J. Phys. B **10**, 1 (1977); A. H. MacDonald and S. H. Vosko, *A Relativistic Density Functional Formalism*, J. Phys. C **12**, 2977 (1979); P. Strange, *Relativistic Quantum Mechanics*, Cambridge University Press, Cambridge (1998), Ch. 9.3; E. Lipparini, *Modern Many-Particle Physics*, World Scientific (2008), Ch. 1.4 – 1.6; E. Engel, R. M. Dreizler, *Density Functional Theory, an Advanced Course*, Springer, Heidelberg (2011), Ch. 8.
26